\documentclass[iop]{emulateapj}

\usepackage{rotating}
\usepackage{amsmath}

\newcommand{\beq}{\begin{equation}}
\newcommand{\eeq}{\end{equation}}
\def\earth{\hbox{$\oplus$}}
\def\arcmin{\hbox{$^\prime$}}
\def\arcsec{\hbox{$^{\prime\prime}$}}

\def\deg{\hbox{$^\circ$}}
\newcommand{\lsim}{\ \raise
-2.truept\hbox{\rlap{\hbox{$\sim$}}\raise5.truept\hbox{$<$}\ }}
\newcommand{\gsim}{\ \raise
-2.truept\hbox{\rlap{\hbox{$\sim$}}\raise5.truept\hbox{$>$}\ }}
\newcommand{\simsim}{\ \raise
-2.truept\hbox{\rlap{\hbox{$\sim$}}\raise5.truept\hbox{$\sim$}\ }}

\renewcommand{\arraystretch}{1.2}

\shorttitle{Structure, Dynamics and SFR of the ONC}
\shortauthors{Da Rio et al. 2014}

\begin{document}

\title{The Structure, Dynamics and Star Formation Rate of the Orion Nebula Cluster}

\author{Nicola~Da Rio$^1$, Jonathan C. Tan$^{1,2}$, Karl~Jaehnig$^1$}
\affil{$^1$Department of Astronomy, University of Florida, Gainesville, FL 32611, USA.\\
$^2$Department of Physics, University of Florida, Gainesville, FL 32611, USA.}
\email{ndario@ufl.edu}

%%=============================================================================================
%%=============================================================================================
%%=============================================================================================

\begin{abstract}
The spatial morphology and dynamical status of a young, still-forming
stellar cluster provide valuable clues on the conditions during the
star formation event and the processes that regulated it.  We analyze
the Orion Nebula Cluster (ONC), utilizing the latest censuses of its
stellar content and membership estimates over a large wavelength
range. We determine the center of mass of the ONC, and study the
radial dependence of angular substructure. The core appears rounder
and smoother than the outskirts, consistent with a higher degree of
dynamical processing. At larger distances the departure from circular
symmetry is mostly driven by the elongation of the system, with very
little additional substructure, indicating a somewhat evolved spatial
morphology or an expanding halo. We determine the mass density profile
of the cluster, which is well fitted by a power law that is slightly
steeper than a singular isothermal sphere. Together with the ISM
density, estimated from average stellar extinction, the mass content
of the ONC is insufficient by a factor $\sim 1.8$ to reproduce the
observed velocity dispersion from virialized motions, in agreement
with previous assessments that the ONC is moderately supervirial. This
may indicate recent gas dispersal. Based on the latest estimates for
the age spread in the system and our density profiles, we find that,
at the half-mass radius, 90\% of the stellar population formed within
$\sim 5$--$8$ free-fall times ($t_{\rm ff}$). This implies a star
formation efficiency per $t_{\rm ff}$ of $\epsilon_{\rm ff}\sim
0.04$--$0.07$, i.e., relatively slow and inefficient star formation
rates during star cluster formation.
\end{abstract}

\keywords{stars: formation, pre-main sequence, kinematics and dynamics; open clusters and associations: individual (Orion Nebula Cluster)  }

%%=============================================================================================
%%=============================================================================================
%%=============================================================================================

\section{Introduction}
\label{section:introduction}

The majority of stars, perhaps including our Sun, have their origin in
clusters \citep{ladalada2003,gutermuth2009}. Thus understanding the
formation of star clusters is important both for their role as the
basic building blocks of galactic stellar populations and for being
the birth environments of most planetary systems.

In spite of this importance, some basic questions about star cluster
formation are still debated, including whether it is dynamically fast
\citep{elmegreen2000,elmegreen2007,hartmann2007} or slow
(\citealt{tan2006} [TKM06]; \citealt{nakamura2007,nakamura2014})
process: in essence, is the duration of star cluster formation similar
to a dynamical time (similar to the free-fall time) of the natal gas
clump or is it much longer? The latter scenario would be consistent
with some theoretical expectations of relatively low star formation
efficiency per free-fall time in turbulent and/or magnetized gas
\citep{krumholz2005,padoan2011} and with turbulence being maintained
by self-regulated protostellar outflow feedback
\citep{nakamura2007,nakamura2014}.% (NL07, NL14).

As discussed by TKM06, there are several ways to try and distinguish
between these scenarios, including considering the morphologies of gas
clumps, the morphologies of embedded stars, assessing the momentum
flux of protostellar outflows, looking at the age spreads of pre-main
sequence stars and the ages of dynamical ejection events. By
considering these factors, especially in the context of the relatively
nearby (414~pc, \citealt{menten2007}) and massive ($\sim$ few thousand
$M_\odot$, \citealt{hillenbrand-hartmann1998}) Orion Nebula Cluster
(ONC), TKM06 concluded the duration of star cluster formation, as
defined by $t_{\rm form,90}$, the time to form 90\% of a cluster's
stars, was $\geq 4 \bar{t}_{\rm dyn}\simeq 8 \bar{t}_{\rm ff}$, where
$t_{\rm dyn}=R/\sigma$, $R$ being the local radius, $\sigma$ being the
1D velocity dispersion and the bar indicating this is the
mass-weighted average of $t_{\rm dyn}$ over the region containing the
stars that count towards $t_{\rm form,90}$. The free-fall time is
defined via $t_{\rm ff} \equiv [3 \pi / (32 G \rho)]^{1/2}$, where $\rho$ is
the volume density, and for a virialized cloud with $\alpha_{\rm vir}
\equiv 5 \sigma^2 R/(GM)\sim 1$ \citep{bertoldi1992}, we have $t_{\rm
  ff}\sim 0.5 t_{\rm dyn}$. For the ONC, TKM06 adopted
$M=4600\:M_\odot$, $\Sigma= M/(\pi R^2)=0.12\:{\rm g\:cm^{-2}}$, $R=1.60$~pc, so
that $t_{\rm dyn}=7\times 10^5$~yr and $t_{\rm ff}=3.5\times
10^5$~yr. They assumed star formation, which is still on-going, has a
duration $t_{\rm form,90}\geq 3$~Myr.

The estimate of $t_{\rm form,90}$ in the ONC is measured most directly
via age spreads of pre-main sequence stars, as revealed by spreads of
luminosity in the HR diagram in comparison with stellar evolutionary
models. However, other factors can also lead to this luminosity
spread, including the difficulty in assigning stellar parameters to
individual stars \citep{dario2010b,reggiani2011} and, episodic
protostellar accretion
\citep{baraffe2009,baraffe2012,hosokawa2011}. \citet{dario2014}
examined the problem of age spread in the Orion Nebula Cluster (ONC)
and concluded, from independent constraints that there is an intrinsic
age spread of $\sim 1.34$~Myr as defined by 1$\sigma$ dispersion in
ages. Assuming a log-normal age distribution or a constant SFR in time,
this leads to $t_{\rm form,90}\geq 4.1, 4.2$~Myr, respectively.

In this paper we re-visit questions of the timescale of star cluster
formation as exemplified by the ONC. First we consider the spatial
structure of stars in the cluster to investigate the TKM06 assertion
of progressively smoother stellar distributions (smaller amounts of
substructure) as a cluster ages. We do this by examining angular
substructure in annuli as a function of radius (adapting methods used
by \citealt{gutermuth2005}). The theoretical expectation is that
center of the cluster, being dynamically older because of its shorter
local dynamical time but similar age spread \citep{dario2012} has a
smoother distribution of stars, since any initial substrucure has had
more dynamical timescales to be erased. This analysis first requires a
careful assessment of the location of the center of mass of the ONC;
and then analysis of radial variation of angular substructure. This is
presented in \S\ref{section:structure}.

Then in \S\ref{section:dynamics}, utilizing the latest mass and age
estimates of the stars in the ONC and allowing for contributions of
gas to the total mass, we examine the mass density profile of the
ONC. Using literature measurements for the velocity dispersion we
refine previous assessment concerning the ONC dynamical equilibrium.
In \S\ref{section:fftime} we compute the ratio of $t_{\rm form,90}$ with $t_{\rm ff}$ as a
function of radius in the cluster. Together with the latest estimate
of the age spread in the system, this allows us to measure the star
formation efficiency per free-fall time, $\epsilon_{\rm ff} = 0.9
\epsilon_* t_{\rm ff}/t_{\rm form,90}$, where $\epsilon_* \equiv
M_*/M_{\rm tot} = \rho_*/\rho_{\rm tot}$. We are able to measure
$\epsilon_{\rm ff}$ both locally as a function of projected radius
(making the simplifying approximation that projected radius is 3D
radius) and as an average interior to a given radius.

\section{The stellar catalogs}
\label{section:catalog}
We assemble catalogs of stellar positions and properties in the ONC from the literature.

We first compiled a sample of all sources with available stellar
parameters, including spectroscopically determined $T_{\rm eff}$ and
$\log L$, as well as (model dependent) ages and masses. In this
context, the H-R diagram of \citet{dario2012} represents the latest
update; it is obtained combining spectral types from either
spectroscopy or narrow-band photometry with optical multi-band
photometry to measure the reddening towards each source and calculate
bolometric luminosities. This sample covers a field of view of about
$\sim40$\arcmin$\times40$\arcmin\ on the Orion Nebula centered
south-west of the Trapezium, and is nearly complete for $A_V<5$~mag
down to the H-burning limit, while also extending into the substellar
regime. Sources flagged as probable non-member contaminants in
\citet{dario2012} have been excluded from the catalog. We further
extended this catalog adding new spectral types from
\citet{hillenbrad2013}; for these stars the extinction, $A_V$, and thus
$\log L$, has been assigned using $BVI$ photometry from
\citet{dario2010a} and adopting the same analysis technique as in that
work. Also, since \citet{dario2012} was incomplete towards the massive
end of the population, due to saturation of their photometry, we added
all of such missing sources adopting the stellar parameters from the
\citet{hillenbrand97} catalog. Last, the masses of the Trapezium stars
have been adjusted to account for their multiplicity, using estimates
of masses for each multiple system from \citet{Grellmann2013}. The
final catalog of optically determined stellar parameters contains 1597
sources.

We also constructed a catalog of near-infrared (NIR) photometry. We
based this on the $JHK$ catalog from \citet{robberto2010}, which
covers an area slightly larger than that of our optical photometry,
and has a very deep detection limit ($3\sigma$ detection at
$J=19.5$~mag, $H=18$~mag), reaching down to planetary masses, and
nearly complete for $A_V<20$~mag for stellar objects. We complement
this sample for the saturated bright end using data from the {\it Two
  Micron All Sky Survey} (2MASS, \citealt{skrutskie2006}). We also
collected the mid-infrared Spitzer survey in Orion from
\citet{megeath2012}. This catalog provides stellar fluxes from the $J$
band to 24~$\rm \mu m$, and includes the classification of sources showing
infrared excess from dusty stellar surroundings (either disks or
protostellar objects). This was based on the multi wavelength, near-
and mid-infrared criteria described in \citet{gutermuth2009}. Last, we
used the X-ray source catalog of \citet{getman2005a} from the {\rm
  Chandra Orion Ultradeep Project} (COUP), and rejected sources
flagged by \citet{getman2005b} as non members of the Orion population
(nebular shocks, extragalactic sources, or unconfirmed members).

All these catalogs have been cross-matched, and the resulting dataset
limited to the sky area covered by the optical data. The X-ray sample
alone has a smaller coverage than the other datasets, complete up to
$\sim 0.11$\deg\ from the Trapezium; all the other catalogs extend to a
radius of $\sim 0.2$\deg\ at any position angle. This corresponds to a
projected distance of $\sim 1.4$~pc from the center, assuming a
distance for the ONC of 414~pc \citep{menten2007}.

Despite the richness of this dataset, it is difficult to precisely
isolate the ONC population, as all these samples alone suffer from
some combination of incompleteness or contamination from
non-members. The optical catalog of stellar parameters is naturally
limited by dust extinction, which is also not spatially uniform. The
NIR photometry is nearly complete at high extinctions, lacking only a
minor fraction of young members in the heavily embedded OMC-1 cloud,
as well as protostellar objects in the vicinity of the Kleinmann-Low
(KL) nebula; however, this sample suffers significant contamination
from Galactic field populations, which increases towards low stellar
luminosities. The X-ray sample is not limited significantly by
extinction, but suffers incompleteness at low stellar masses
($M<0.2$~M$_\odot$) and in the Trapezium cluster due confusion from
the broad tails of the PSF of the bright objects. Last, the Spitzer
survey also suffers from incompleteness in the detection at low
masses, and confusion in the core due to the lower angular resolution
compared to the other samples.

In the following sections, we will assume the combination of the
optical parameter and the X-ray sample as representative of the
spatial structure of the ONC. This joint sample, although
somewhat
incomplete, is virtually immune from field contamination and not
biased by patchy extinction. The remaining sources in the IR samples
will be considered when assessing the total stellar mass and its
radial density profile.

\section{The Structure of the ONC}
\label{section:structure}

\subsection{The Center of the ONC}
\label{subsection:ONC_center}

\begin{figure*}
\includegraphics[width=0.32\textwidth]{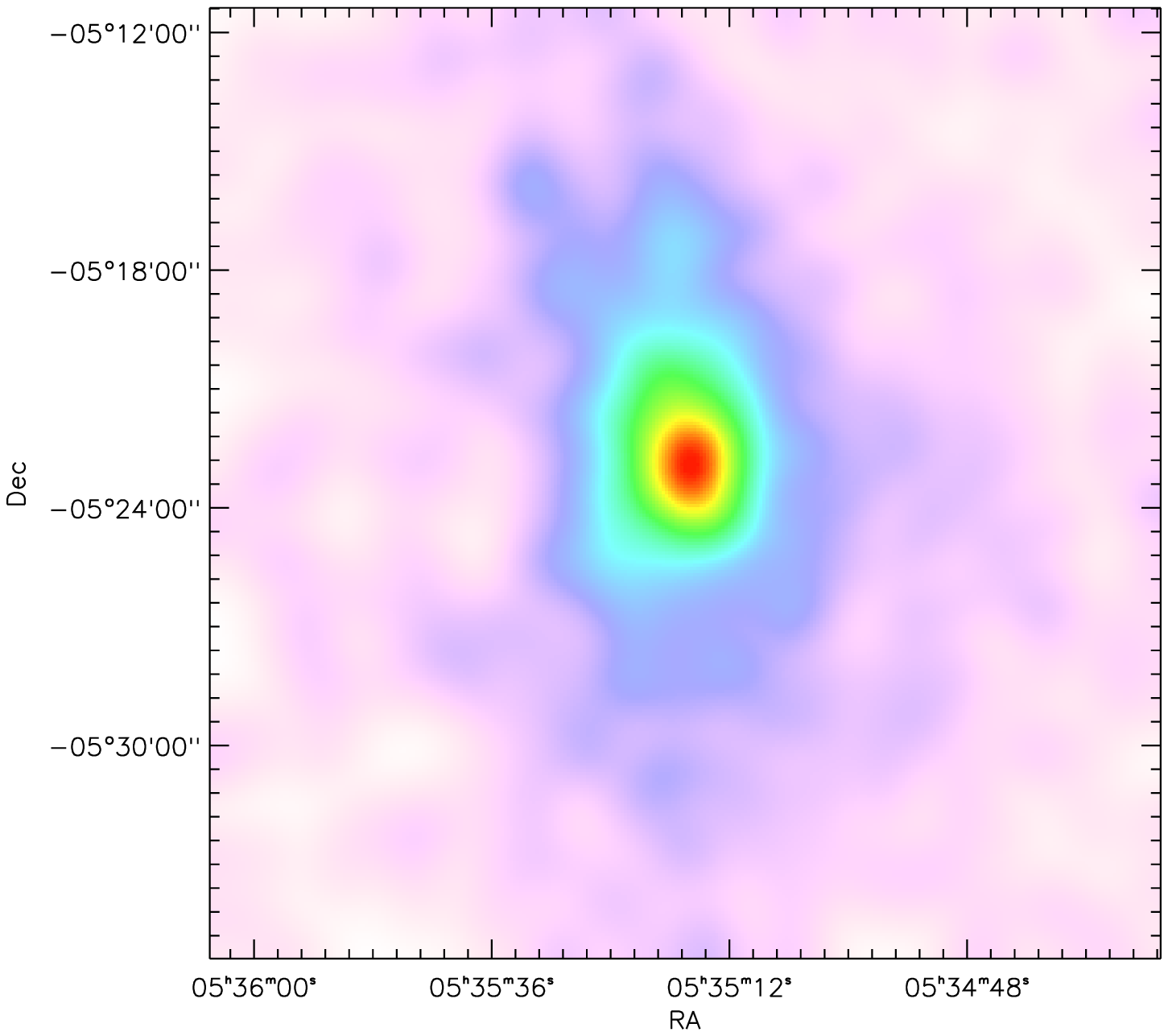}
\includegraphics[height=0.67\textwidth, angle =90 ]{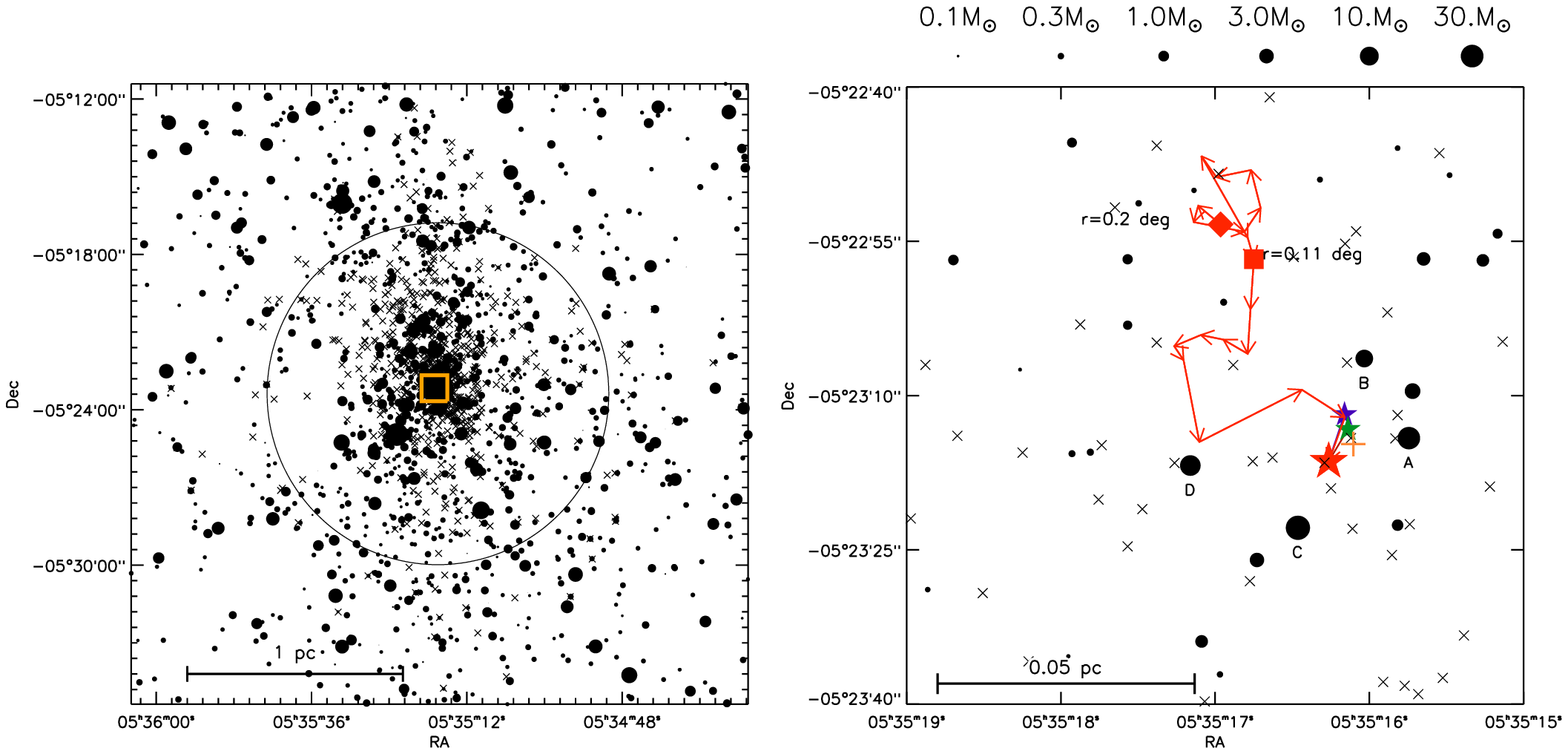}
\caption{
\emph{Left panel}: ONC stellar density map, limited to an angular box
of 0.4\deg\ (2.9~pc) centered on $\theta^1C$ for our sample of optical
and X-ray members. \emph{Middle panel:} The catalog of positions;
symbols denote stellar mass if known, crosses indicate X-ray sources
with unknown mass.
The circle delimits the aperture of radius
0.11\deg\ (0.9~pc) fully contained within the COUP X-ray field of
view. The orange box delimits the area shown in the Right
panel. \emph{Right panel:} The Trapezium region; the red arrows
indicate the moving center of mass of the population decreasing the
considered aperture from the entire catalog (red diamond) to the COUP FOV
aperture (red square) to the central radius 0.01\deg\ (0.072~pc) (red
star). Blue and green stars
mark, respectively, the shift of the final result when excluding
$\theta^1$C from the sample, or including it but as if it were located at the
position of the ejection event proposed by \citet{chatterjee2012}
(orange plus symbol). \label{figure:center}}
\end{figure*}

\begin{figure*}
\includegraphics[width=0.32\textwidth]{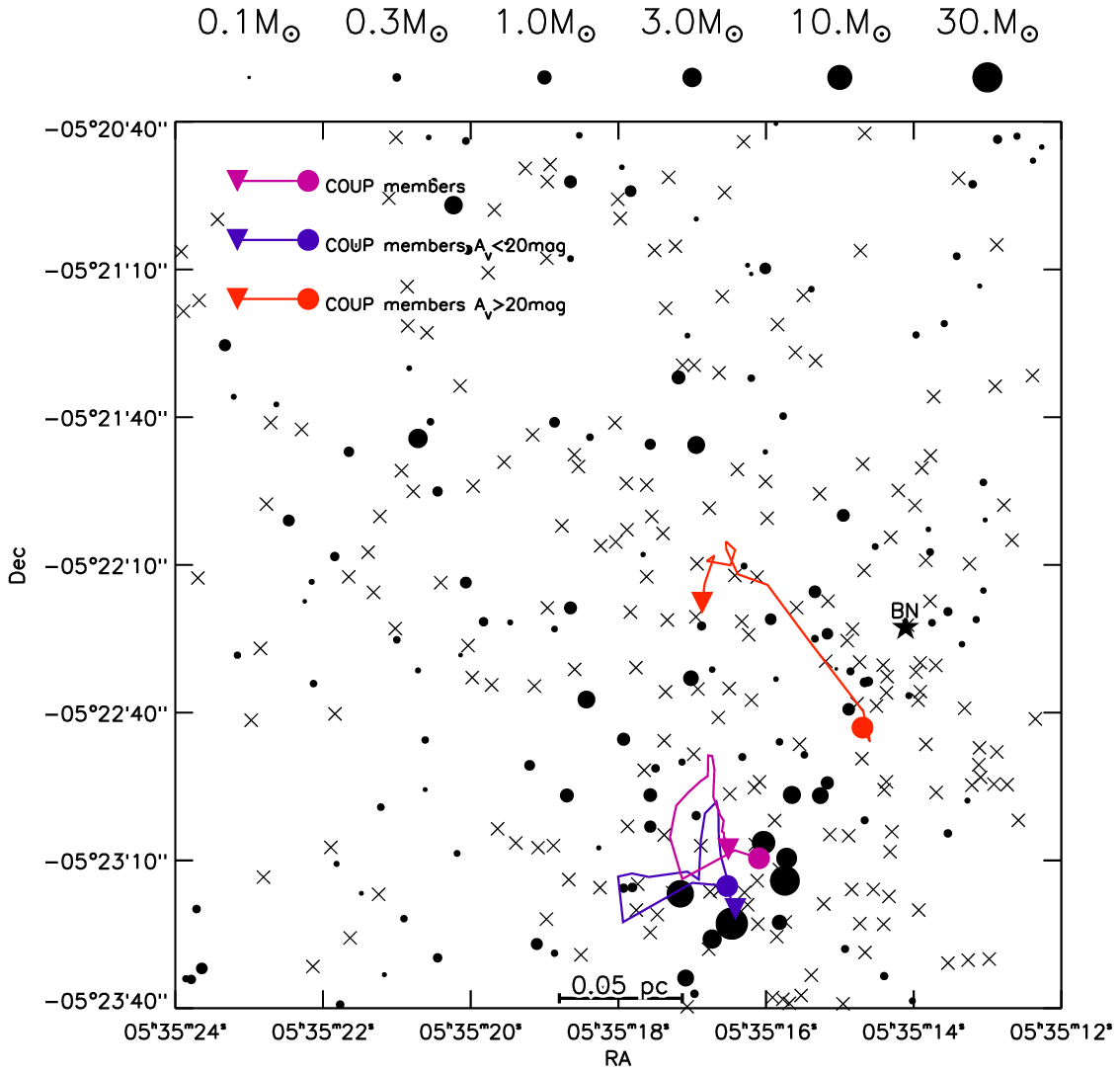}
\includegraphics[width=0.32\textwidth]{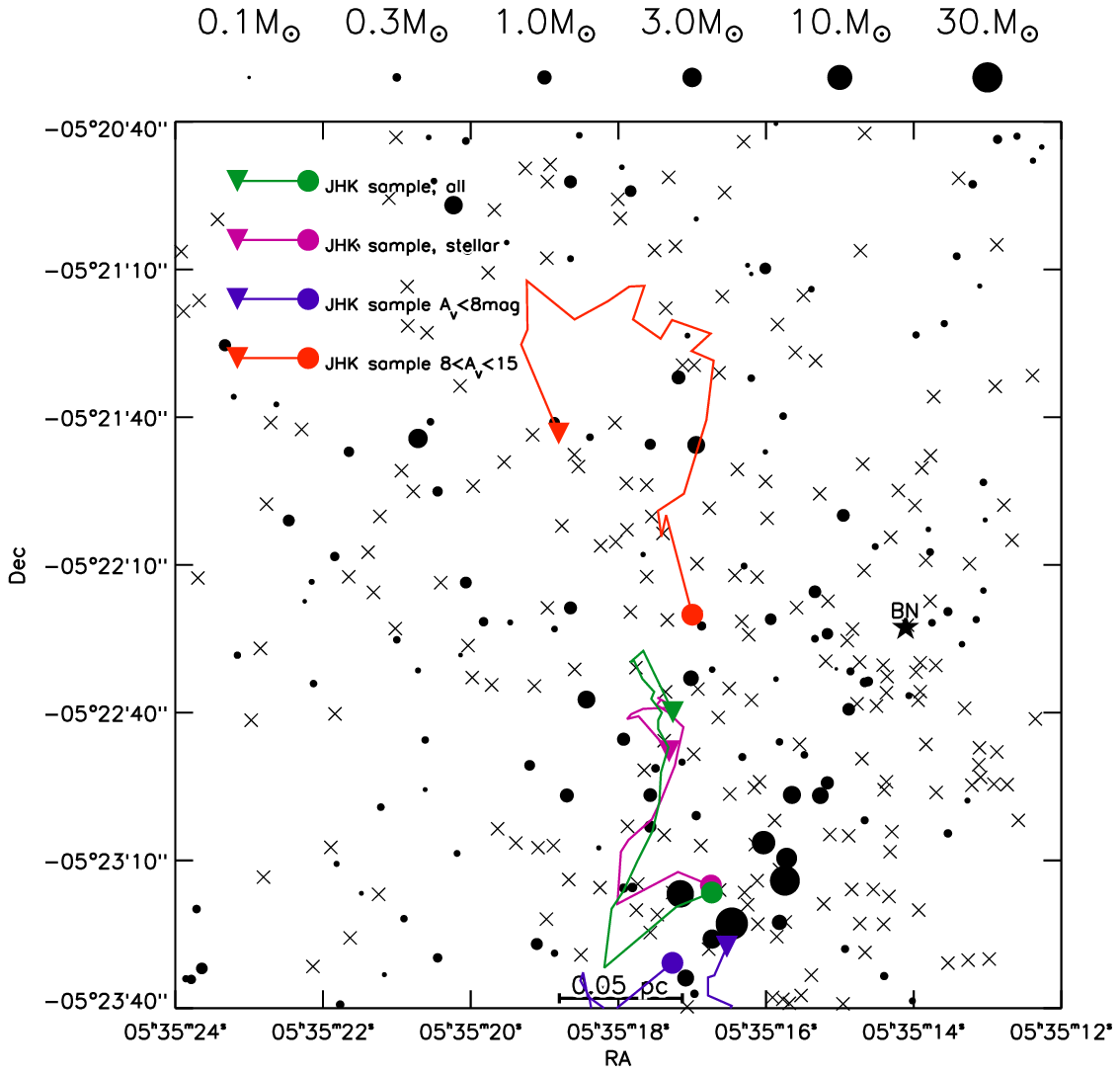}
\includegraphics[width=0.32\textwidth]{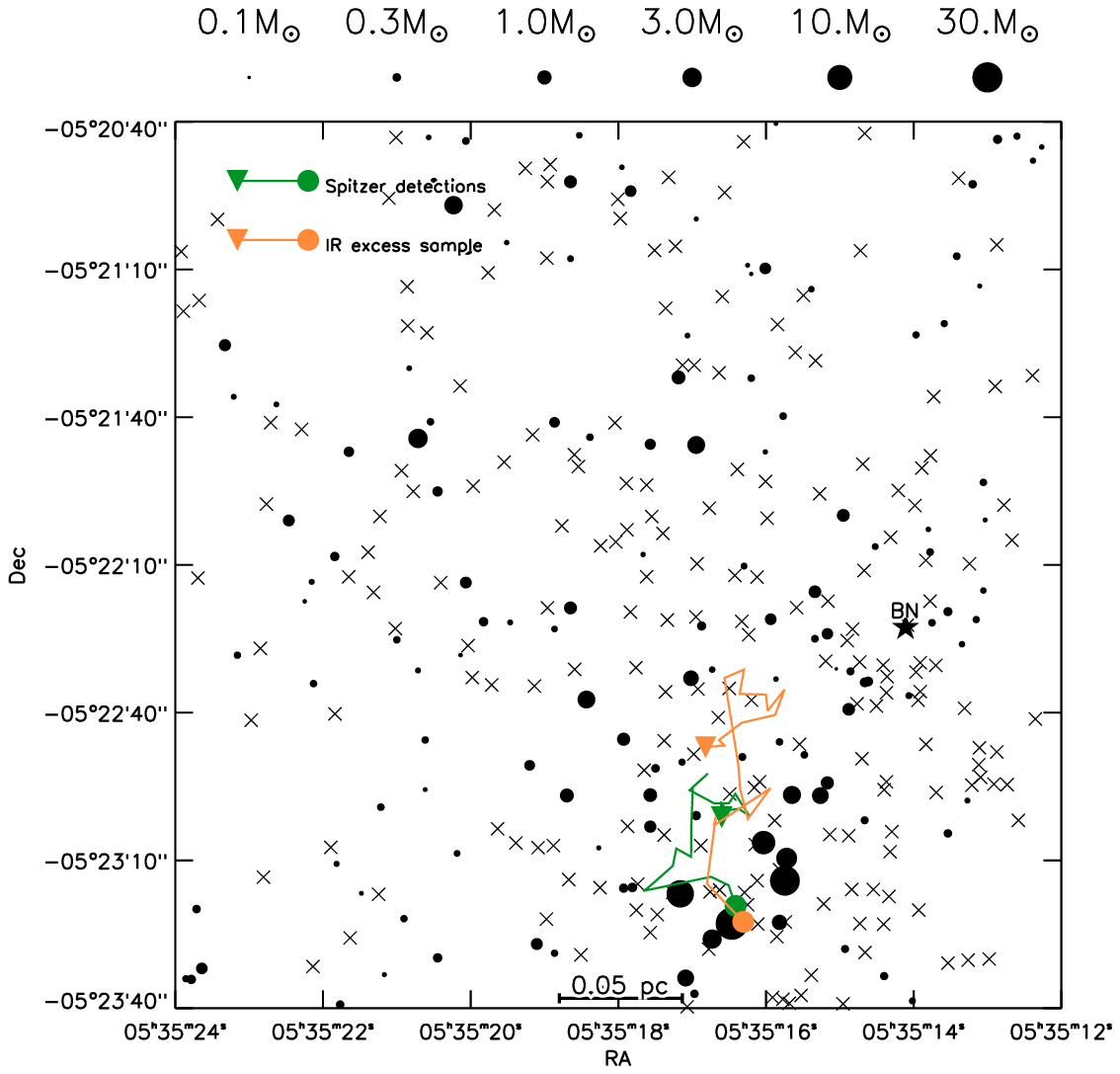}
\epsscale{1}
\caption{
Same as Figure \ref{figure:center}, but tracing the displacement of
ONC center of mass assuming different samples. \emph{Left panel}: the COUP X-ray
sample, and the same considering separately lightly and heavily
embedded sources. \emph{Middle panel}: the NIR photometry, considering
all detections, detections above the CMD threshold dividing stellar
members from BD and contaminants, and the latter divided in 2
extinction samples. \emph{Right panel}: the Spitzer detections sample,
and the same limited to sources showing infrared excess from their
SED. In every panel the triangles denote the ``starting'' positions from the largest aperture, and the circles the final centers at small scales.\label{figure:plots_center_changing_catalogs}}
\end{figure*}

It is well established that the massive stars of the Trapezium cluster lie
roughly at the center of the stellar population of the ONC; here we
aim to better characterize the actual position of the center of mass
of the population. \citet{hillenbrand-hartmann1998}, based on ellipse
fitting on isophotes on an optical+NIR stellar density map found the
center of the stellar positions to be located $\sim 25$\arcsec\ north
of $\theta^1$C and slightly to the west, just outside the Trapezium
cluster. \citet{feigelson2005}, based on the COUP X-ray sample, noted
that the heavily embedded sources ($\log N_{\rm H}>22$~cm$^{-2}$,
corresponding to $A_V\gtrsim 6$~mag) are systematically offset to the
north-west of the Trapezium compared to the lightly obscured
members. This, however, may be in part due to the spatial variations
of extinction, which are present in the ONC at all column densities.

We have considered the merged catalog of available optical stellar
parameters with X-ray members to evaluate the position of the center
of mass of the ONC.  The collected census counts 2228 members in
total, or 1901 sources within the square area of size
0.4\deg\ centered on $\theta^1$C common to all catalogs, to which we
restrict our analysis. About 2/3 of the stars in this sample have a
mass estimate from optical studies; the remaining are X-ray members
with no mass estimate. We assign to these sources a mass
$M=0.5M_\odot$, i.e., the mean mass of the \citet{kroupa2001} initial
mass function. In fact, as mentioned, the X-ray sample is incomplete
below 0.2$M_\odot$, so its mean stellar mass should be higher; on the
other hand all the massive stars in the ONC have optical parameters,
so this lowers the mean mass of the X-ray sample with no available
mass estimate. The stellar positions included in our final catalog of
optical and X-ray sources are shown in Figure \ref{figure:center},
left and center panels; the circle of radius 0.11$\deg$ in the center
panel delimits the maximum aperture fully contained in the X-ray field
of view.

The center of the ONC has been computed in an iterative way, from
large to small scales: first we considered the largest circular
aperture contained in the whole area shown in Figure
\ref{figure:center} and determined its center of mass. This has been
then used as a center of a slightly smaller circular aperture, where
the sample has been limited to, in order to re-derive its center of
mass. The procedure has been iterated to progressively smaller
circles, down to $\sim$0.6\arcmin, each time using as the aperture
center the center of mass of the previous one. Figure
\ref{figure:center}, right panel shows the displacement of the center
of mass from largest area, to the X-ray complete aperture and then to
the smallest aperture. We find that at large apertures the center of
mass is displaced $\sim30$\arcsec\ (0.05~pc) north of the $\theta^1$C
and outside the Trapezium, in agreement with previous works. When we
reduce the aperture to sample the center at smaller scales, this
progressively moves inside the Trapezium, indicating some degree of
asymmetry at different scales. We will consider the latter center,
located at $\alpha_{\rm J2000}=05^h~35^m~16.26^s$; $\delta_{\rm
  J2000}=-05\deg~23$\arcmin$~16.4$\arcsec~ as our bona-fide center of
the ONC.

We have tested how this result is sensitive to the assumed value of
stellar mass for uncharacterized X-ray sources, and find no
displacement (less than 1\arcsec) if this value is lowered to $M_{\rm X-ray}=0.3\:M_{\odot}$.

Finally, we note that the local ONC center of mass lies close to the
point where, tracking proper motions in reverse, as proposed by
\citet{chatterjee2012}, $\theta^1$C and the Becklin-Neugebauer (BN)
object were co-located $\sim 4500$ years ago. It is proposed that a
strong gravitational slingshot interaction ejected BN into the
molecular cloud at $\sim 30\:{\rm km\:s^{-1}}$ with $\theta^1$C
recoiling in the opposite direction to its current location. Thus, we
have recalculated the ONC center of mass both removing $\theta^1$C
from the sample, or displacing its position to the point of the past
interaction. In both cases, as shown in Figure \ref{figure:center},
our calculated center of mass moves even closer to the interaction
point. This qualitative argument could strengthen the hypothesis of
\citet{chatterjee2012}, as the system of $\theta^1$C, the most massive
star in the cluster, would tend to settle in the very center of the
cluster via gentle interactions with other ONC members. Its current
displacement from the ONC center of mass is then explained as a result
of its strong interaction with BN. If true, this scenario has the
potential to place constraints on the formation time and radial
displacement from cluster center of the formation site of $\theta^1$C.

\subsection{Displacement of the ONC Center}
\label{subsection:ONC_center_changing_catalog}

We now study how the center of the ONC, and its variations upon
aperture scale, depends on the sample of sources we have adopted. First
we aim to test if the incompleteness of our optical and X-ray sample
could bias our derived ONC center; second, we look for systematic
displacements of the center of the population as a function of dust
extinction, which provides some indication of the distribution of
stars and gas along the line of
sight. \citet{hillenbrand-hartmann1998} found no significant
variations of the spatial distribution of sources observed in the
optical and in the near infrared. Using X-ray derived extinctions,
which reach to higher column densities than near infrared photometry,
\citet{feigelson2005} however found the embedded population to be more
concentrated to the east of the Trapezium cluster, with
over-concentrations aroung the BN/KL region and the OMC-1S
region. \citet{feigelson2005}, however, set the division between the
lightly and heavily obscured samples at $\log N_{\rm H}=22$~cm$^{-2}$
which corresponds to $A_V\sim6$~mag, a value low enough to be
sensitive to the non uniformity of the average extinction of the ONC
(see, e.g., the extinction maps from \citealt{scandariato2011}), rather
than a value separating highly embedded members which cannot be
detected in the optical or NIR.

We first considered the X-ray sample alone, and converted the column
densities $\log N_{\rm H}$ derived from the X-ray spectral analysis of
\citet{getman2005a} into dust extinction $A_V$ assuming
the relation $N_{\rm H}/A_V=1.58\times10^{21}$~cm$^{-2}$ \citep{vuong2003} . We then separated the sample in 2 sub samples with
$A_V<20$~mag and $>20$~mag, and computed the center. This was
performed as in \S\ref{subsection:ONC_center}.
Figure \ref{figure:plots_center_changing_catalogs} (left panel) shows
that the center of the ONC remains confined inside the Trapezium when
considering the sample with up to 20 mag of extinction, which
corresponds to $\sim85\%$ of the population. The remaining small
fraction of heavily embedded sources remains spatially centered
$\lesssim0.1$~pc to the northwest of the Trapezium in the direction of
the BN/KL region. This is not surprising as the KL region is
associated to the densest molecular core within the OMC-1 filament
\citep{johnstone1999,grosso2005}.

Next, we have computed the center of the ONC adopting the NIR
photometric sample. We have separately considered either the entire
sample, and that restricted to sources brighter than the reddening
vector in the $J$ versus $J-H$ diagram corresponding to a mass of
$0.075 M_\odot$. \citet{robberto2010} has shown that below this locus
in the NIR CMD, roughly corresponding to the substellar mass range,
the vast majority of sources are faint field contaminants rather than
young brown dwarfs. Figure
\ref{figure:plots_center_changing_catalogs} (middle panel) shows removing or not
this faint population of probable contaminants has little effect on
the derived center of the ONC. As for the X-ray sample, we divide the
NIR catalog according to extinction. This was roughly computed by
deredding the $J$ vs $J-H$ CMD on a 2.5~Myr \citet{siess2000}
isochrone. Again, the center of the lightly embedded population is
confined within the Trapezium cluster, but this moves north at higher
$A_V$, at a higher declination than for the X-ray embedded
population. This result has two origins: first, the NIR sample is severely
limited by dust extinction in the densest BN-KL and OMC-1S regions,
whose combined embedded populations (dominated by OMC-1S) would thus
shift the center of the obscured population to the south-west; second,
at high extinction the NIR catalog from \citet{robberto2010} has a
spatially variable completeness, and all sources with $A_V>15$ are
located on a stripe around $\delta=-5$\deg18\arcmin, which coincides
with the overlapping area between adjacent exposures of the imaging
mosaic, and thus has higher effective photometric depth. Such a feature
in the spatial completeness therefore biases the center towards the
north at high $A_V$.

Last, we considered the Spitzer sample; Figure
\ref{figure:plots_center_changing_catalogs} (right panel)
%{subsection:ONC_center_changing_catalog}
shows that its center remains compatible with that of the other
catalogs, and does not vary when limiting to sources showing infrared
excess emission. This suggests that there are no significant spatial offsets between young members with disks and diskless sources, although their radial profile might not be identical.

\subsection{Ellipticity}
\label{subsection:ellipticity}

\begin{figure}
\plotone{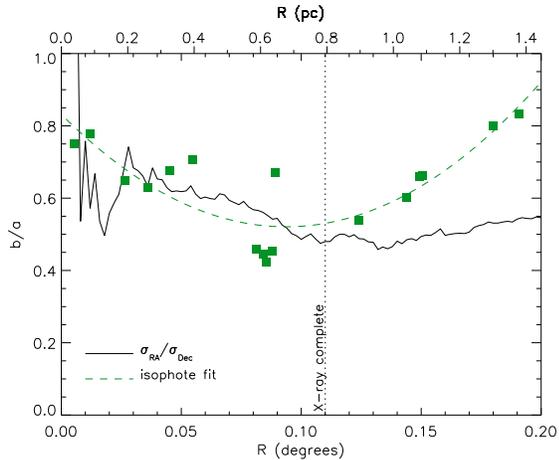}
\caption{
Axes ratio, $b/a$, of ONC stars. The green squares are the best
elliptical fits to isophotes on the ONC map, and trace the local
ellipticity. The green dashed line
line is a 2$^{nd}$ order polynomial fit to these results. Black solid
line: cumulative flattening of the population in the north-south
direction as a function of maximum distance in declination from the
center, estimated from the ratio of the positional dispersions in RA
and Dec (see text).
\label{figure:b_over_a}}
\end{figure}

The ONC population is known to be elongated in a direction close to
north-south, which follows the local filamentary structure of the
Orion~A molecular cloud \citep{johnstone1999,muench2008}.
\citet{hillenbrand-hartmann1998} fitted ellipses to isophotes on a
stellar spatial density map in the ONC, obtaining average ellipticity
$e=1-b/a=0.30$, identical for their optical and NIR sample,
with a tilt of the major axis at about 10\deg\ counterclockwise from
the north-south direction.

We remeasure the ellipticity of the ONC considering our catalog of
optical parameters and X-ray members as representative of the
structure of the cluster. We evaluate both a ``local'' and a scale
dependent ``overall'' ellipticity. We compute the first in a similar
way as in \citet{hillenbrand-hartmann1998}, we generate a stellar
density map, and smooth it with a Gaussian kernel of
36\arcsec\ ($\sim0.08$~pc). We recompute the tilt of the major axis of
the cluster, which from our catalog is only 7\deg, and choose to
neglect it as negligible throughout our analysis. We also constrain
the center of each fitted ellipse to be our reference ONC center,
computed on the same catalog, derived in \S\ref{subsection:ONC_center}.

We also compute a overall ellipticity of the entire ONC population
within varying distances from the center. To this end, we proceed in
an iterative way: we consider square boxes centered on our ONC center,
and compute the ratio between the standard deviations in RA and Dec of
the sources contained. If the ratio departs from one, we change the
size of the box in RA (generally diminishing it given the north-south
elongation of the ONC), and recompute the ratio of the dispersions of
the sample contained in this rectangular area. We iterate until the
measured ratio of sample standard deviation converges to the ratio of
the sides of the box.

The result, in terms of axis ratio $b/a=1-e$ as a function
of distance in declination from the center is shown in Figure
\ref{figure:b_over_a}. Our isophote fits show that the core of the
cluster is relatively round, and the distribution becomes more
elliptical at increasing radii, reaching a value $b/a\sim 0.5$; this
is compatible with \citet{hillenbrand-hartmann1998}'s results, which
were obtained up to a distance (semimajor axis) of 0.14\deg\ from the
center. At larger distances, however, the cluster becomes again
rounder, with a semimajor axes ratio of $\sim0.8$. The black line in
Figure \ref{figure:b_over_a} shows the cumulative $b/a$ as a function
of radius: its increase at distances from the center
($\gtrsim0.12$\deg) is more modest since the
enclosed stellar content remains dominated by stars in the more
elliptical region.

\subsection{Angular Substructure}
\label{subsection:dispersion}

\subsubsection{The Angular Dispersion Parameter}
\label{subsection:dispersion-method}

\begin{figure*}
\plotone{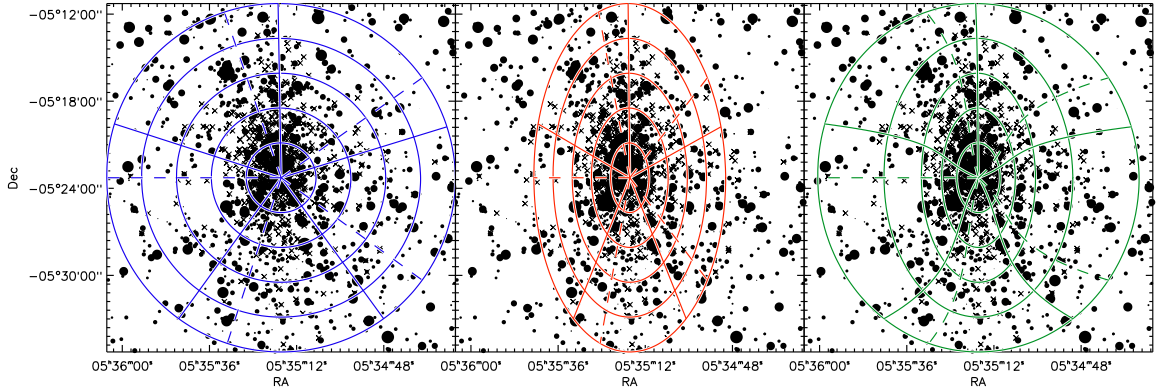}
\caption{
Example of the subdivision of the ONC population in sectors and
annuli, assuming circular symmetry (left panel, blue lines), constant
ellipticity (center panel, red lines), or radially variable
ellipticity (right panel, green lines). The dashed lines show another
possible orientation of the sector
pattern. \label{figure:plots_sectors}}
\end{figure*}

\begin{figure}
\plotone{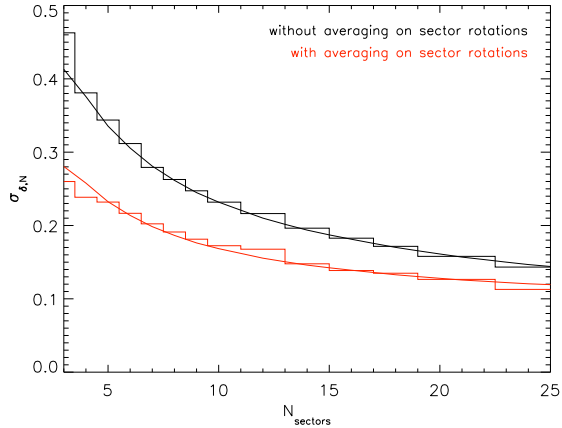}
\caption{
The measured error, $\sigma_{\delta,N}$, in the angular dispersion
parameter, $\delta_{\rm ADP, N}$, versus number of sectors, $N$, from
our Monte Carlo simulation, assuming a fixed pattern of sectors for
each simulated stellar distribution (black histogram) or averaging
$\delta_{\rm ADP, N}$ for the $n_{\rm tot}$ possible rotations of the patters (red
histogram). The black line is the analytic prediction for the first
from Equation \ref{equation:error_in_d}, the red line its scaling to
match the red histogram. \label{figure:error_in_d_vs_sectors}}.
\end{figure}

\begin{figure}
\plotone{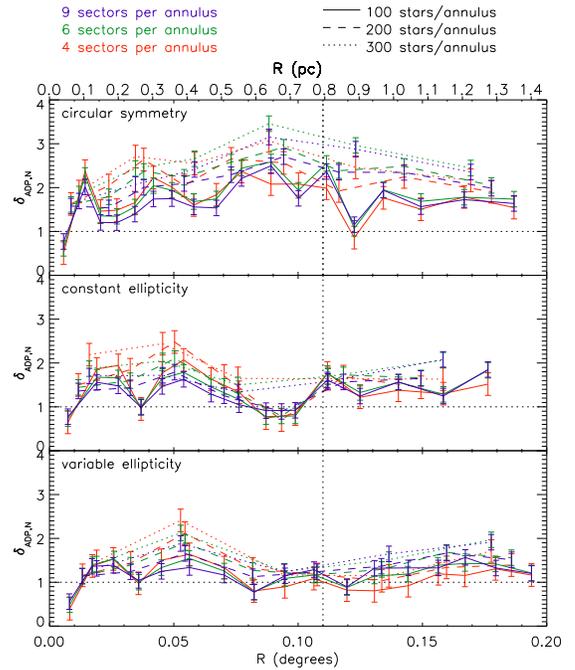}%{plots_pvalue.eps}
\caption{
Angular dispersion of stellar counts in sectors normalized on that
expected for Poisson statistics, as measured by the ADP, as a function
of radial distance from the center of the selected annuli. Lines
corresponds to different numbers sectors and stars per annulus, and
for the assumption of circular symmetry as opposed to elliptical
symmetry. The vertical dotted line at delimits the X-ray complete
sample at $r<0.11$\deg .
 \label{figure:plots_dispersion}}
\end{figure}

We estimate the degree of angular sub-structure in the ONC, or
conversely, the smoothness of the stellar distribution, using a
technique analogous to the \emph{``azimuthal asymmetry parameter''}
(AAP) defined by \citet{gutermuth2005}. This is based on dividing the
spatial stellar distribution in equally sized circular sectors, and
comparing the dispersion of the number counts among different sectors
with the hypothesis of being drawn from a uniformly random
distribution of position angles. Varying the width (or number) of the
sectors allows one to probe positional substructure at different
azimuthal multipole moments. The radial dependence of the degree of
angular substructure can be investigated by further isolating the
population within concentric annuli before counting sample numbers
within azimuthal sectors.

In addition to radial variation, we will also generalize the AAP of
\citet{gutermuth2005} to account for cluster ellipticity. We thus
define a new {\it angular dispersion parameter} (ADP), $\delta_{\rm
  ADP, N} (R)$. For each annulus, the number $n_i$ of stars in each
$i$-th sector is counted over a total of $N$ sectors; the quantity
$\delta_{\rm ADP, N}$ is then defined as follows:
\begin{equation}
\label{equation:deltaADP}
\delta_{\rm ADP, N}= \sqrt{\frac{1}{(N-1) \overline{n}}\sum\limits_{i=1}^{N} \big( n_i - \overline{n})^2}=\sqrt{\frac{\sigma^2}{\sigma^2_{\rm Poisson}}}
\end{equation}
\noindent where $\sigma$ is the standard deviation of the $n_i$
values, $\overline{n}$ is the average of the number of stars per
sector in the considered annulus, and $\sigma_{\rm Poisson}$ is the
expected standard deviation due to Poisson statistics. When the
annuli follow the local or mean elliptical shape of the cluster, we indicate
this via $\delta_{\rm ADP,e, N}$ and $\delta_{\rm ADP,\bar{e},N}$, respectively.
The ADP simply corresponds to the measured sample standard deviation of counts
in sectors normalized on that expected assuming Poisson
statistics. Practically, an azimuthal random distribution of stars
would produce a measured
$\delta_{\rm ADP, N}\sim1$;
in presence of intrinsic cluster sub-structuring, the
measured dispersion increases to values $>1$.

Strictly speaking, since the sample variance $\sigma^2$ follows a
scaled $\chi^2$ distribution, as
$(N-1)\sigma^2/\sigma_{\rm Poisson}^2$ follows a $\chi^2_{N-1}$ distribution with $N-1$ degrees of freedom, we have that the expected value of $\delta_{\rm ADP, N}^2$ is 1 if the stellar distribution is azimuthally random, but the non linearity of the square root in Equation \ref{equation:deltaADP} lowers the mean of $\delta_{\rm ADP, N}^2$ to $\sim0.93$ for $N=4$, to $\sim0.95$ for $N=6$ and to 1 for $N\rightarrow\infty$.

Since $\delta_{\rm ADP, N}$ is a random variable subject to a
statistical error, deviations from 1 are expected even for a random
distribution of stars. From the relations mentioned above, the standard error in $\delta_{\rm ADP, N}$
will be:
\begin{equation}
\label{equation:error_in_d}
\sigma_{\delta,N} = \sqrt{{\rm Var}[\delta_{\rm ADP, N}]}=\sqrt{{\rm Var}\left[\frac{1}{N-1}\chi^2_{N-1}\right]},
\end{equation}
\noindent which does not depend on the number of stars, but on the
number of sectors. For a small number of sectors, this error is
relatively large: for example $\sigma_{\delta,N}\simeq 0.25$ for 4 sectors and
$\simeq0.17$ for 10 sectors.

A way to lower this error is to decrease the probability of the
measured $\delta_{\rm ADP, N}$ deviates from the expected value
because of the particular orientation of the sector pattern (e.g., the
edge between two contiguous sectors oriented in the north-south
direction as shown in Figure \ref{figure:plots_sectors}). Instead, the
value of $\delta_{\rm ADP, N}$ can be computed for multiple
orientations of the sector pattern and the results averaged. The total
number of unique redistributions of $n_{\rm tot}=N \bar{n}$ sources
within an annulus among the different sectors obtained by rigid
rotations of the sector patter is $n_{\rm tot}$; so we compute
$\delta_{\rm ADP, N}$ for each of all these cases and average the $n_{\rm tot}$
results.

We characterize the decrease of $\sigma_{\delta,N}$ due to our
averaging process through Monte Carlo simulations. We generate a large
number of simulated stellar distributions with random positions within
a circular aperture, and estimate the error $\sigma_{\delta,N}$ as the
standard deviation of the dispersions measured for each
realization. We repeat the experiment changing the the number of stars
within the aperture (from 20 to 5000) and the number of sectors. Also,
in each case we separately test the cases of measuring $\delta_{\rm
  ADP, N}$ assuming a fixed pattern of sectors, or performing an
average over the $n_{\rm tot}$ possible orientations of the patterns
for each simulated distribution. Results show that $\sigma_{\delta,N}$
is independent of $n_{\rm tot}$ also in the latter case. Instead (see
Figure \ref{figure:error_in_d_vs_sectors}), the value of
$\sigma_{\delta,N}$ when the rotational averaging is performed is
lower than that for a fixed sector pattern by an amount that depends
on the number of sectors, being from $\sim30\%$ smaller for 4 sectors
and $\sim20\%$ for more than 20 sectors. In our analysis that follows,
we will always adopt the value of $\delta_{\rm ADP, N}$ obtained by
averaging over sectors rotations, and the error bar $\sigma_{\delta,N}$ on
these derived from our Monte Carlo simulations (red line in Figure
\ref{figure:plots_sectors}).

We have established that the statistical uncertainty in $\delta_{\rm
  ADP, N}$ does not depend on the number of stars, $n_{\rm tot}$, in
an annulus or circular aperture. However, if a population of stars is
not azimuthally uniform, a given degree of substructure will lead to
different values $\delta_{\rm ADP, N}>1$ depending on $n_{\rm
  tot}$. This is because of the normalization of $\delta_{\rm ADP, N}$
over the expected dispersion for a Poisson distribution: increasing
$n_{\rm tot}$ causes the relative expected standard deviation of the
star counts in sectors (over the total counts) to decrease. Thus, for
a given relative increase in the measured standard deviation of counts
produced by substructure, the higher the number of stars, the higher
the value of $\delta_{\rm ADP, N}$. Thus, when comparing the measured
$\delta_{\rm ADP, N}$ between different star clusters, or different
radial bins for the same cluster, or for different assumed samples for
the same cluster, the number of sources within an aperture or annulus
must be fixed.

Before we analyze the properties of the ADP, $\delta_{\rm ADP, N}$, in
the ONC, we briefly characterize the typical ranges of the variation
of this parameter between very smooth and highly substructured stellar
populations, and in particular consider Globular Clusters (GCs) and
the PMS stars in Taurus-Auriga. For the GCs, we adopt the catalogs
from the \emph{ACS Survey of Galactic Globular Clusters}
\citep{sarajedini2007,anderson2008}, which includes HST photometry of
about 50 GCs. For the Taurus association, we use the census of PMS
stars from \citet{kenyon2008} which includes 383 members. For each GC
and for Taurus we derive the center of the cluster as in \S\ref{subsection:ONC_center}, divide the population in circular annuli
and sectors, forcing a fixed number of sources within each annulus,
and compute $\delta_{\rm ADP, N}$. Then we average the results for
multiple annuli in Taurus and all annuli of all GCs. Table
\ref{table:taurus-ONC-GCs_disp} shows the results, compared with that
obtained in the ONC assuming the optical parameters + X-ray sample,
and imposing either 50 or 100 stars per annulus. Such a low number,
very small compared to the number of sources in the GCs and also in
the ONC, is required to allow a meaningful comparison with the small
sample of the Taurus region.
\begin{table}[t]
\centering
\caption{Average measured dispersion $\delta_{\rm ADP, N}$}
\label{table:taurus-ONC-GCs_disp}
\begin{tabular}{l|r|r|r}
\multicolumn{4}{c}{50 stars per annulus} \\ [0.5ex]
  \hline
    & 4 sectors & 6 sectors & 9 sectors \\
  Globular Clusters & 0.92 & 0.95  & 0.96 \\
  ONC & 1.39 & 1.40 & 1.31 \\
  Taurus & 2.77 & 3.01 & 3.03 \\ [0.5ex]
  \hline
  \multicolumn{4}{c}{ } \\
   \multicolumn{4}{c}{100 stars per annulus} \\ [0.5ex]
  \hline
    & 4 sectors & 6 sectors & 9 sectors \\
  Globular Clusters & 0.92 & 0.95  & 0.97 \\
  ONC & 1.80 & 1.79 & 1.63 \\
  Taurus & 2.98 & 3.52 & 3.54 \\ [0.5ex]
  \hline
\end{tabular}
\end{table}

Table \ref{table:taurus-ONC-GCs_disp} shows a trend clear in the ADP
from the smooth, dynamically old, GCs, where the $\delta_{\rm ADP,
  N}\simeq1$ indicates a near random azimuthal distribution of sources
(indeed such low values are expected due to the
regularization imposed by the global potential of the cluster, while
increased values may result from a spread in stellar masses), to the
ONC where departures from Poisson smoothness are detected ($d\simeq1.4$ for
50 stars per annulus, $\delta_{\rm ADP, N}\simeq1.8$ for 100 stars per
annulus), to the substructured distribution in Taurus leading to an
azimuthal dispersion up to twice as large as in the ONC.  These
results highlight the ability of the azimuthal dispersion parameter
$\delta_{\rm ADP, N}$ to trace small departures from angular spatial
smoothness: another technique such as the minimum spanning tree $Q$
parameter \citep{cartwright2004} is a powerful tracer of substructure
for clumpy spatial distributions, but in the ONC ($Q\sim0.8$) would
merely indicate central concentration.

\subsubsection{Radial Dependence of the Angular Dispersion Parameter in the ONC}
\label{subsection:dispersion-ONCradial}

We now look for radial variations in the ADP of the ONC,
i.e. $\delta_{\rm ADP, N}(r)$. Also, we account for the ellipticity
of the cluster and consider separately 3 assumptions
\begin{description}
  \item[Circular symmetry] we simply divide the stellar sample in
    concentric circular annuli in RA and Dec to derive $\delta_{\rm
      ADP, N}(r)$.
  \item[Constant ellipticity] we assume elliptical annuli, with an
    axes ratio $b/a=0.55$ corresponding to the overall ellipticity we
    have determined within large apertures ($a>0.1$\deg) from the ONC
    center (see Figure \ref{figure:b_over_a}) to derive $\delta_{\rm
      ADP,\bar{e},N}(r)$. The position angles of the segments separating
    neighboring sectors are corrected to maintain equal areas within
    each sector.
  \item[Variable ellipticity] we assume the polynomial fit to the best
    fit isophotes shown in Figure \ref{figure:b_over_a}, allowing the
    flattening of subsequent annuli to vary with the distance from the
    center, to derive $\delta_{\rm ADP,e,N}(r)$. As in the previous
    case, the edge between neighboring sectors is defined to force the
    area of all sectors in each annulus to be constant, this produces
    curved lines separating sectors.
\end{description}
An example of the 3 configurations we explore is shown in Figure
\ref{figure:plots_sectors}.  As before, we assume the combined sample
of optical and X-ray sources.
% with optical parameters and X-ray membership.

Figure \ref{figure:plots_dispersion} reports the radial dependence of
$\delta_{\rm ADP, N}$ from our ONC stellar sample, for multiple
configurations of number of sectors and stars per annulus, as reported
in the legend. As anticipated, increasing the number of stars per
annulus leads, on average, to a larger measured value of $\delta_{\rm
  ADP, N}$, whereas we do not detect significant differences in the
results changing the number of sectors, i.e., the angular mode of
substructure, from $N=4$ to 9. Assuming circular symmetry leads to the
highest dispersion, with a broad peak at $\sim0.9$\deg from the
center. This corresponds to the distance of maximum ellipticity of the
cluster (see Figure \ref{figure:b_over_a}); such a feature in the
dispersion profile disappears when accounting for ellipticity (lower
panels of Figure \ref{figure:plots_dispersion}). This shows that the
radial increase in the angular dispersion assuming circular symmetry
is not indicative of clumpy substructure, but is largely due to the
elongation of the ONC. Allowing for elliptical sectors with varying
ellipticity leads to the lowest dispersion, almost radially constant
at an average value $\delta_{\rm ADP, N}\sim1.2$--$1.6$.

\begin{figure*}
\center
\epsscale{0.75}
\plotone{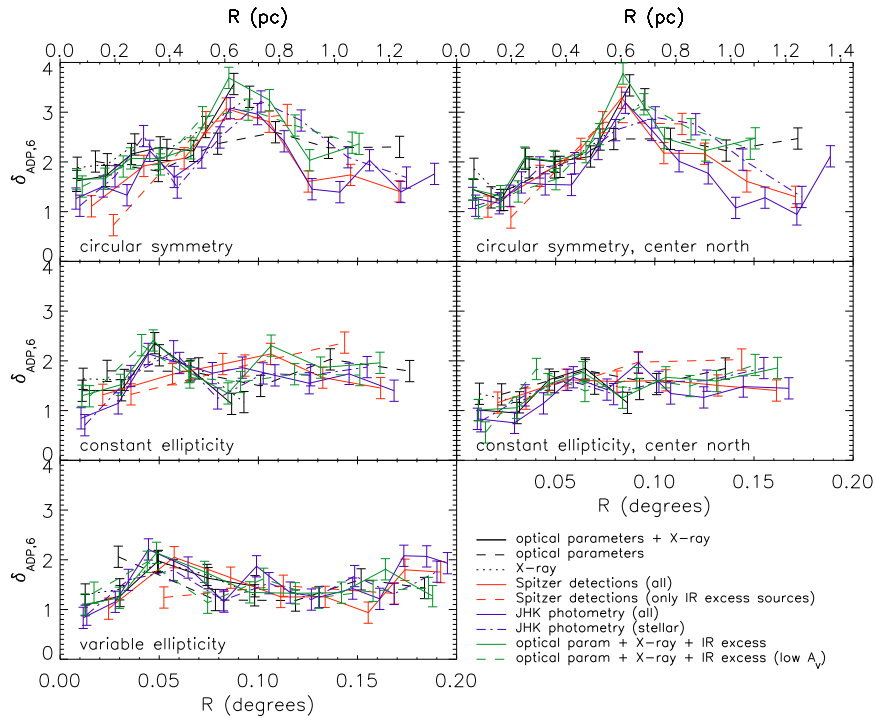}
\epsscale{1}
\caption{
Dispersion versus distance from the center for different selections of
the stellar samples, and method to account for ellipticity as
indicated in the legend. Left panels denote values of $d$ for annuli
centered on our ONC bona-fide center of mass, right panels assume the
center of mass within an aperture of $r=0.11$\deg\ (see Figure
\ref{figure:center}), to the north of the Trapezium. The value of $d$
is computed defining annuli to contain 200 stars and assuming 6
sectors \label{figure:dispersion_changing_sample}}
\end{figure*}

Since, as mentioned in \S\ref{section:catalog}, our
representative sample of sources with optical parameters or X-ray
membership remains somewhat incomplete at the very low-mass end of the
IMF, and at projected distances from the center $>0.11$\deg\ where
part of the region has not been observed in X-rays, we have tested the
radial behavior of $\delta_{\rm ADP, N}$ measured from different
assumptions for the stellar catalog. In particular, we have compared
the cases of assuming the sample with either optical parameters or
X-ray membership separately; we have considered the entire $JHK$
photometry and that restricted to the stellar luminosity range,
largely immune to contamination which is predominat at fainter
luminosities. We have also considered the catalog of Spitzer detections from \citet{megeath2012}, and separately its
subsample of IR excess sources. These results are shown in the
left column of Figure \ref{figure:dispersion_changing_sample}, for the
three assumptions regarding the cluster ellipticity. We find that the
qualitative behavior of $\delta_{\rm ADP, N}$ is largely unaffected by
the choice of the sample, with typical differences between the
measurements comparable with the statistical errors affecting each
value. This indicates that the degree of substructure we detect in the
ONC is not being set by residual contamination, incompleteness of
the adopted stellar sample,
or patchy extinction.

Since in \S\ref{subsection:ONC_center} we have shown that the
center of mass of the ONC shifts when computed within apertures of
different radii, and at large scales is $\sim 20$\arcsec\ north of the
Trapezium (Figure \ref{figure:center}), we have also assessed if the
radial trend of $\delta_{\rm ADP, N}$, and its absolute values are
driven by this effect. Figure \ref{figure:dispersion_changing_sample},
right column, reports the radial dependence of $\delta_{\rm ADP, N}$
when computed centering the pattern of annuli to the center of mass of
the ONC within an aperture of 0.11\deg\ from the Trapezium (red square
symbol in Figure \ref{figure:center}). Also in this case, we find
little or no change compared to the radial trend of $\delta_{\rm ADP,
  N}$ from our final selected center of the cluster.

The very center of the ONC appears to have a significantly smaller
angular dispersion parameter, $<1$, with the values rising by factors
of a few by $R=0.05$\deg. This behavior is independent of whether or
not ellipticity is allowed for.

Moreover, as we have shown in \S\ref{subsection:ellipticity},
the inner part of the ONC is rounder than at larger distances. Such
behavior is expected, considering that core has a shorter dynamical
timescale than the halo, thus stellar interactions can smooth out the
spatial distributions faster.

For the outer regions, once ellipticity is allowed for, then the level
angular substructure is relatively constant with radius. This indicates that the peak in $\delta_{\rm ADP,
  N}$ measured at $R\simeq0.6$~pc assuming circular symmetry is mainly driven by the elongation of the system, which is highest at this distance from the center (Figure \ref{figure:b_over_a}).
  On close inspection to our catalogs, we also noted that at intermediate distances increase in  $\delta_{\rm ADP,
  N}$ is also influenced by the relatively underabundance of sources to the east of the Trapezium compared to the west, an asymmetry already noted by \citet{feigelson2005}.
The value of  $\delta_{\rm ADP,
  N}$ after correcting for ellipticity, thus, is
larger than the mean value in the dynamically old globular clusters,
but smaller than in the more dispersed Taurus region. This may
indicate there has been some dynamical processing if the stars in the
ONC formed with the same initial substructuring as
Taurus. Alternatively, if the stars in this extended region are part
of an expanding halo of weakly bound or unbound cluster members, which
formed in a more central location, then this could also explain the
observed flattening of $\delta_{\rm ADP,N}$ beyond $\sim 0.4$~pc.

N-body simulations utilizing varying initial conditions have
investigated the temporal evolution of the $Q$ parameter, the stellar
surface density distribution and mass segregation
\citep[e.g.,][]{allison2009,allison2010,parker2014}. Analysis needs to
be extended to the $\delta_{\rm ADP, N}$ parameter, which could
further constrain the initial conditions and dynamical evolution
of the ONC.

Further observational data is also needed. In a forthcoming paper, we
will study $\delta_{\rm ADP, N}$ and its radial dependence in a large
sample of young clusters, spanning a range of masses, densities and
ages.

\section{Dynamics of the ONC}
\label{section:dynamics}

In this section we utilize our collected datasets of the ONC to
reevaluate its dynamical status. In particular, we constrain
the overall mass profile, both due to stellar and gas potential, and
compare it with kinematic studies from the literature to asses the
virial equilibrium of the system.

\subsection{Stellar density profile}
\label{subsection:profile_stars}
\begin{figure}
\plotone{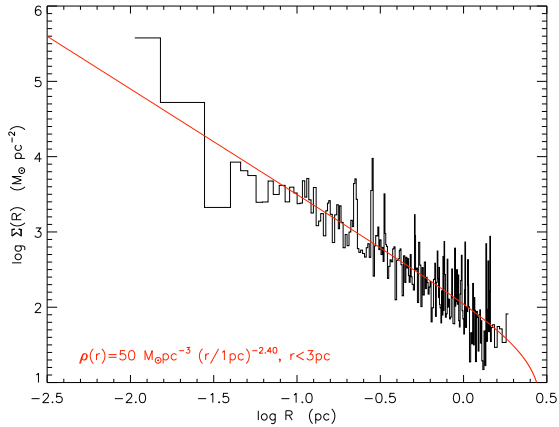}
\caption{
Measured mass surface density profile of the ONC, adopting the sample of
sources with available stellar parameters or X-ray membership (black
line). The red line shows the best fit of a truncated 3D power law
profile.
\label{figure:profile}}
\end{figure}

Following the work of \citet{hillenbrand-hartmann1998}, we study the
radial density profile of the stellar population in ONC. For
simplicity here we neglect the ellipticity of the cluster and derive
an average radial profile, adopting circular symmetry in the plane of
sky. We consider all the stellar catalogs described in
\S\ref{section:catalog}; as anticipated, these have been restricted to
a square area with a size of 0.4\deg\ (2.9~pc) centered on our
bona-fide ONC center of mass (\S\ref{figure:center}), where we have
full coverage for the optical, near infrared and Spitzer catalogs. The
maximum distance from the center to the corners of this area is thus
0.28\deg\ or $\sim$2~pc.  We divided the samples in radial annuli,
each containing 10 stars, up to the maximum distance from the center,
and measured the projected stellar mass density summing the stellar
masses in each annulus and dividing the result by the area of each
annulus. As in \S\ref{subsection:dispersion}, stars with no
available mass estimate had been assigned a mass of 0.5~$M_\odot$. The
results for the outer annuli, which are in part outside our square
field of view, have been corrected to account for this incompleteness.

The mass surface density profile of the ONC has been computed for
different combinations of the stellar samples. In particular, we
considered the full photometric samples (optical, near infrared and
Spitzer), the sub-samples of sources with optically derived
parameters, the youth tracers (IR excess, X-ray emission), and several
combinations of these selection criteria. An example of these surface
stellar density profiles is shown in Figure \ref{figure:profile}, for
the sample of sources with either optically derived parameters or
X-ray membership.

Unlike in \citet{hillenbrand-hartmann1998}, our mass surface density
profiles $\Sigma(R)$ tend not show a flattening in the core regions,
but appear to follow roughly a straight line in logarithmic axes in
the entire radial range. Thus, instead of adopting King models, we
assume that the 3D stellar density $\rho(r)$ follows a power-law
profile up to a maximum radius $r_{\rm max}$ which sets the boundary
of the cluster:
\[
 \rho(r) =
  \begin{cases}
   \rho_0\  \bigg(\displaystyle\frac{r}{\rm 1pc}\bigg)^{-\alpha} & \text{if } r \leq r_{\rm max} \\
   0        & \text{if } r > r_{\rm max}
  \end{cases}
\]
We assume $r_{\rm max}=3$~pc, and varying $\alpha$ we project the 3D
$\rho(r)$ in 2D for $\Sigma(R)$; these latter are then fit to the data
using a $\chi^2$ minimization, thus determining the best fit $\alpha$
and $\rho_0$.

\begin{deluxetable*}{lcccc}
\tablecolumns{5}
%\tablewidth{0pt}
 \tablecaption{ONC population density profile parameters
 \label{table:profile_param}}
 \tablehead{
 \colhead{\ }       & \colhead{\ }       & \colhead{\ }       & \colhead{\ }       & \colhead{\ }       \\[2pt]
  \colhead{}       & \colhead{\ }                     & \colhead{\ }  & \colhead{\ } & \colhead{\ }                 \\
  \colhead{Sample} & \colhead{$\rho_0$}             & \colhead{$\alpha$}          & \colhead{$N_*$ $(<2$~pc)}               & \colhead{$M_*$ ($<2$~pc)}   \\
  \colhead{\ }       & \colhead{$(M_{\odot}{\rm pc}^{-3})$} &\colhead{ }           &\colhead{ }                & \colhead{$(M_\odot)$}      \\
  \colhead{ }       & \colhead{ }                     & \colhead{ }          & \colhead{ }               & \colhead{ }                  }
 \startdata
 \hline \\[5pt]
 a) all sources                                  & 105 & 2.05 & 4323 & $2676$ \\[1pt]
 b) $JHK$ photometry                             & 95 & 1.98 & 4129 & $2386$ \\[1pt]
 c) $JHK_{\rm stellar}$                          & 55 & 2.25 & 2745 & $1539$ \\[3pt]
\hline \\[3pt]
 d) optical photometry                           & 52 & 1.90 & 2969 & 1285 \\[1pt]
 e) optical parameters                           & 33 & 2.05 & 1927 & 846  \\[1pt]
  \phm{blahblah} f) young                        & 17 & 2.24 & 1009 & 465 \\[1pt]
  \phm{blahblah} g) old                          & 17 & 1.88 & 939 & 405 \\[3pt]
 \hline \\[3pt]
 h) optical param + X-ray                        & 50 & 2.40 & 2595 & 1591 \\[1pt]
 i) optical param + X-ray + IR excess            & 55 & 2.35 & 2741 & 1667 \\[1pt]
 j) optical param + IR excess                    & 46 & 2.02 & 2238 & 1156 \\[1pt]
 k) IR excess                                    & 23 & 2.01 & 1094 & 578 \\[1pt]
\hline \\[3pt]
l) optical param + X-ray +  $JHK_{\rm stellar}$  & & & & \\[1pt]
                     \phm{blahblah}$A_V<1$~mag & 18 & 1.96 & 915 & 452\\[1pt]
                     \phm{blahblah}$A_V<3$~mag & 32 & 2.29 & 1621 & 916\\[1pt]
                     \phm{blahblah}$A_V<10$~mag & 50 & 2.26 & 2733 & 1421\\[1pt]
                     \phm{blahblah}$A_V<30$~mag & 66 & 2.24 & 3132 & 1850\\[1pt]
                     \phm{blahblah}$A_V<100$~mag & 66 & 2.27 & 3230 & 1886\\[3pt]
\hline \\[3pt]
 \enddata

 \tablecomments{Samples are defined as follows. $a)$: any individual source detected in the optical photometry of \citet{dario2010a}, the NIR photometry of \citet{robberto2010} complemented with 2MASS, Spitzer and X-ray members from \citet{getman2005b};
 $b)$: NIR photometry catalog;
 $c)$: as b) but excluding sources in the CMD zone populated by brown dwarfs and contaminants (see \citealt{robberto2010});
 $d)$: optical photometry from \citet{dario2010a};
 $e)$: sample of optically derived stellar parameters from \citet{dario2012};
 $f)$ and
 $g)$: sources with available age estimate from the HRD, divided as younger or older than the mean cluster age.
 $k)$: Spitzer detection showing evidence of IR excess from circumstellar material \citep{megeath2012}. From
 $h)$ to $l)$: combination of the above criteria. }
\end{deluxetable*}

Table \ref{table:profile_param} reports the best-fit density profile
parameters, as well as the extrapolated number of sources and total
stellar mass within 2~pc from the center. Overall, the power law
exponent $\alpha$ is found to be close to that of single
isothermal sphere ($\alpha_{\rm SIS}=2$), and only weakly affected by the
criteria for selecting the stellar sample, despite a factor of several
difference in the number of stars.  The values $\alpha\gtrsim2.3$
obtained for samples that include the X-ray sources is in part biased
towards steeper slopes by the incomplete X-ray coverage at large
distances from the center. Conversely, the optical photometric catalog
shows a flatter slope than other samples, possibly due to lower
completeness in the central regions of the ONC because of the bright
nebular background in the vicinity the Trapezium. Similarly, stars
with isochronal ages older than the mean cluster age are more likely
to be missed in the central regions, as they are fainter than younger
sources for the same mass. Lastly, the exponent from the fit to the
entire $JHK$ photometry, which include prominent background
contamination at substellar luminosities, turns out to be flatter than
that obtained restricting to the stellar mass range, where
contamination is minimal. This is expected as the contamination from
Galactic field sources is not as centrally concentrated as the ONC
members. From all these comparisons, we estimate a bona fide value
$\alpha\simeq 2.2$ for the ONC population.
We emphasize that in principle a power law density profile is unphysical, in that it has infinite density at $r=0$. However, for $\alpha<3$ the mass contribution of the core does not diverge, and for the isothermal case each radial bin in linear units contributes the same amount of mass. Since our measurements find no deviation from a single power law down to $r\sim10^{-2}$~pc, changing the model to remove the singularity at smaller radii would have no effect on our analysis.

The assessment of the real value of $\rho_0$ from the values listed in
Table \ref{table:profile_param} is critical to constrain the actual
stellar mass of the ONC, and requires some considerations.  First the
value $\rho_0\simeq 100\: M_{\odot}{\rm pc}^{-3}$ determined for
entire sample of unique detections summed from each catalog, as well
as the whole $JHK$ sample, is significantly overestimated due to the
large contamination at faint luminosities. The normalization value,
hence the total mass of the cluster nearly halves when restricting to
near infrared sources above the stellar mass threshold of
\citet{robberto2010}, below which nearly each source is not a
member. Even this sample, however, suffers from some contamination and
some incompleteness. For example, within the same field of view and
luminosity range ($M\gtrsim0.2\:M_\odot$), $\sim15\%$ of NIR sources
are not X-ray members; this increases to $\sim 25\%$ in the whole
stellar luminosity range; this is both due to increasing
incompleteness of the X-ray survey, and increasing contamination
towards lower luminosities. However, the X-ray sample reaches deeper
extinctions than our JHK photometry.  Within an aperture of
0.11\deg\ from the ONC center, we find 25\% of X-ray sources with no
counterpart in the $JHK$ stellar sample. Thus, incompleteness and
contamination should roughly cancel out, in stellar number, in the
stellar luminosity range of the $JHK$ sample. Yet, this sample will be
affected by further incompleteness, at faint luminosities under the
stellar threshold in the $JH$ CMD, and somewhat in the core of the
region, where confusion limits the X-ray sample. This can be noted
from Table \ref{table:profile_param} considering the sample of
optically derived parameters (which extends somewhat in the substellar
regime), X-rays and IR excess sources: this sample is virtually immune
from contamination but likely incomplete, outside the FOV of the X-ray
sample, and at low luminosities near the core. Its measured
normalization constant $\rho_0=55~M_\odot\:{\rm pc}^{-3}$ is identical to
that of the $JHK$ sample at stellar luminosities.

Based on the above data, it is not clear what the degree of residual
incompleteness is in these samples that is caused by substellar
objects, unresolved binaries, and confusion in the center.
With some uncertainty, we thus assume an additional $25\%$ of total
stellar mass. Thus we infer that the ONC stellar population is well
represented by a density distribution:
\beq
\rho_{\rm stars}(r)\simeq\ 70~M_\odot~{\rm pc}^{-3} \bigg(\frac{r}{1{\rm
    pc}}\bigg)^{-2.2},\ r<3{\rm ~pc}.
\label{equation:stellar_profile_fit}
\eeq

We emphasize that this simple model is intended to be representative of the overall dynamical contribution from stellar mass in the ONC; on smaller scales some degree of substructure, as well as elongation, remain present (see \S\ref{section:structure}). Other studies  \citep[e.g.,][]{rivilla2013,kuhn2014} have also analyzed the spatial structure of the ONC population within the inner pc of the region, finding that the stellar distribution is well matched by a superposition of a denser core - basically coincident with the Trapezium - surrounded by a shallower halo.

Last, as we have shown in Section \ref{section:structure}, the heavily embedded population, though it accounts for a small fraction of the population, appears slightly offset from the main cluster, following the densest cores in the region and the integral shape filament within the OMC-1. In this study we do not consider this part of the population as a separate population to be removed from the sample, as we infer it will eventually be one with the rest of the system during the upcoming early evolution of the ONC. However, we check if the spatial distribution of the heavily embedded population affects the surface density profiles we have derived. The last rows of Table \ref{table:profile_param} report the fitted profile parameters for the sample of sources with available $A_V$ (either optical spectroscopy, NIR CMD, or X rays), limited below varying upper limits in extinction. Except for the lowest extinctions, which appear slightly less centrally concentrated, most likely since they belong to the very external shell of the system toward our direction, the power law exponent remains largely unaffected by the chosen cut in extinction.

\subsection{ISM density}
\label{subsection:profile_with_gas}

%%%%%%%%%%%%%%%%%%%%%%%%%%%%%%%%%%%%%%%%%%%%%
\begin{figure}
\plotone{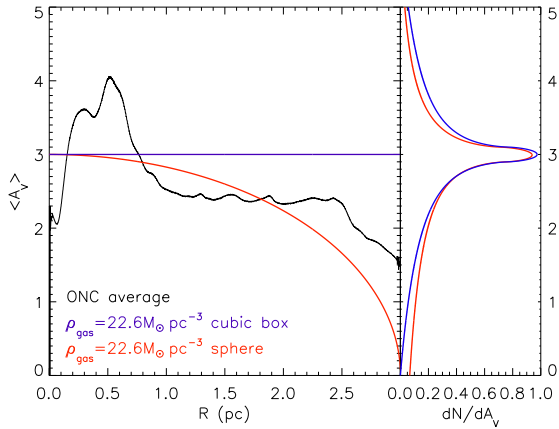}
\caption{
Average extinction affecting the ONC members as a function of
projected radius from the center, obtained from the average ONC
members $A_V$ map from
\citet{scandariato2011}. \label{figure:AV_vs_radius}}
\end{figure}
%%%%%%%%%%%%%%%%%%%%%%%%%%%%%%%%%%%%%%%%%%%%%

\begin{figure}
\plotone{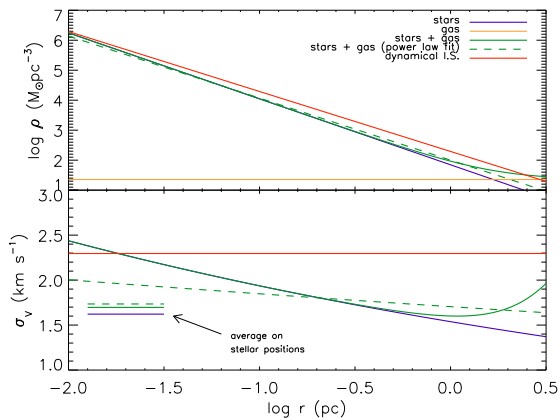}
\caption{
\emph{Top panel:} The estimated volume density profile of the ONC due
to stars, gas, and total (stars + gas), together with the best
power-law fit to the latter as reported in the Legend. The red line
represents the density profile of a singular isothermal sphere needed
under the assumption of dynamical equilibrium given the observed
velocity dispersion in the system
(\S\ref{subsection:equilibrium}). \emph{Bottom panel:} the predicted
1D velocity dispersion, $\sigma_v$ as a function of radius for the these models. The
horizontal lines on the left indicate the overall cluster $\sigma_v$,
computed averaging the curves with a weighting proportional to the fractional stellar mass at each
radius. \label{figure:plots_profile_with_gas}}
\end{figure}

The distribution, density and total mass of the ISM in the ONC is not
well constrained.
Some of the densest regions of the OMC cloud (the KL region and the
OMC-1 South cores) can reach column densities of up to $A_V\sim
100$~mag \citep{bergin1996,scandariato2011,lombardi2011}, but the
total mass estimates of these clumps and cores can be uncertain by at
least a factor of two, given uncertainties in temperatures, dust
emissivities and gas to dust mass ratios. In addition much of the gas
in the region lies somewhat behind the ONC. This is demonstrated by
the large difference between the total ISM column densities integrated
along the line of sight and the relatively small extinction affecting
ONC stellar members which peaks at $A_V=1$--$2$~mag and presents a tail
extending to $A_V\gtrsim 5$--$10$~mag
\citep{hillenbrand97,dario2010a}. As anticipated, only a minor
fraction of young stellar members appear highly obscured.

\citet{scandariato2011} used near-infrared data, together with optical
parameters where available, to derive both the total extinction map --
from statistics of background stars -- as well a map of the average
$A_V$ affecting the ONC members. We have adopted the latter and
computed its mean value as a function of angular distance from the
center of the ONC. The result is shown in Figure
\ref{figure:AV_vs_radius}. The mean stellar $A_V$ is nearly constant
at all distances from the centers, at a value $\langle
A_V\rangle\simeq2.5$--$3$~mag. The peak extiction at $R\simeq0.5$pc is due to the \emph{Dark Bay} \citep{odell2001}, an obscuring cloud in slight foreground with respect to the ONC population and HII region, located north-east of the Trapezium.

If we assume that the ISM is uniformly
distributed also along the line of sight, we can translate this column
density of dust into a total volume density of ISM. Since the $A_V$
distribution is skewed, this is probably not the case, and the ISM
density on average increases moving into the cluster along the line of
sight, reaching high values for the few very embedded
objects. However, this approximation is fair at the midplane of the
system. If we thus assume that the ISM is uniformly distributed in
either a cubic box or a sphere around the cluster center with the
radius of 3~pc, the same truncation we have assumed in Figure
\ref{figure:profile} for the stellar distribution, this translates into
an optical depth $A_V=1~{\rm mag\: pc}^{-1}$ along the line of
sight. Assuming the dust to gas relation from \citet{vuong2003}
$N_{\rm H}/A_V=1.58\times10^{21}~$cm$^{-2}$ and solar abundance of He, this
corresponds to a constant gas density:
\begin{equation}
 \rho_{\rm gas}\simeq22~M_\odot ~{\rm pc}^{-3}.
\label{equation:gas_density}
\end{equation}
\noindent

It could be argue that part of the extinction towards ONC sources may originate from foreground galactic ISM unrelated to the cluster. E.g., \citet{odell2008} finds that the Orion Nebula HII region is obscured by $A_V\sim2$~mag of neutral material. However, the vast majority of such veil remains located within the stellar system, meaning that part of the ONC population is well in front of the HII region. This is confirmed by the fact that spectroscopic measurements of extinctions towards individual ONC members \citep{hillenbrand97,dario2010a,dario2012} measure $A_V$ values as low as $\sim 0$ with no evidence, within the uncertainties, of a positive minimum threshold value. Therefore the foreground extinction from Galactic ISM must be of negligible amount, up to no more than a few tenths of magnitude, negligible compared to the mean ONC extinction shown in Figure \ref{figure:AV_vs_radius}). Given the relatively large uncertainties already present in our mass estimates for both stars and gas, we thus do not attempt to constrain and remove the small foreground extinction.

Figure \ref{figure:AV_vs_radius} also shows, together with the radial
trend of the average $A_V$, the trend expected assuming this
uniform amount of ISM either in a cubic or spherical geometry;
in both cases the simple model is fairly adequate to reproduce the
data.

The $A_V$ map from \citet{scandariato2011} was derived from NIR data,
thus lacking potentially heavily embedded sources that could, although
they are a small fraction of the population, shift the mean stellar
$A_V$ to higher values. We have thus also considered the extinctions
derived from $\log N_{\rm H}$ from the COUP X-ray sample. Indeed, the X-ray
mean $A_V$ is $\sim 15$~mag; this value however is strongly biased by
the asymmetry of the distribution, with a few sources with derived
values exceeding 100~mag, and large relative errors in the
measurements. The median $A_V$ is 3.8~mag, in line with the mean from
Scandariato et al. (2011). Also, the errors in $A_V$ from the X-ray analysis
correlate with $A_V$, and the values of $N_{\rm H}$ appear nearly symmetric
around a mean value of $10^{21.7}$~cm$^{-2}$,
which corresponds to $A_V\simeq3$. Thus we conclude that our estimate
for the value $\rho_{\rm gas}\simeq 22\:M_\odot\: {\rm pc}^{-3}$ adequately
represents the present average ISM content within the ONC.

The above ISM density, when compared to the stellar mass profile (Equation
\ref{equation:stellar_profile_fit}) is quite small: it is smaller than
the stellar density for $r<1.37$~pc, radius containing $73\%$ of the
stellar mass within 2~pc, or $53\%$ within 3pc, and negligible in the
cluster core. If we approximate the contribution of stars and gas to
the radial profile with a power law, the results depends on the
considered range in radii. Limiting to the range of distances spanned
by our stellar density profiles (e.g., Figure \ref{figure:profile}),
the approximate total density follows
\begin{equation}
\rho_{\rm total}\sim 100 \bigg(\frac{r}{1\:{\rm pc}}\bigg)^{-2.07}\:M_\odot \:{\rm pc}^{-3}.
\label{equation:stellar_plus_gas_density}
\end{equation}
We will utilize this as an estimate for the total observed mass, for
comparison with the dynamical mass
(\S\ref{subsection:equilibrium}). Figure
\ref{figure:plots_profile_with_gas} summarises the contribution to the
density profile from equations \ref{equation:stellar_profile_fit} and
\ref{equation:gas_density}, and the approximation to the total of
equation \ref{equation:stellar_plus_gas_density}.

\subsection{Dynamical Equilibrium}
\label{subsection:equilibrium}

It has been pointed out in several works
\citep[e.g.,][]{hillenbrand-hartmann1998,scally2005} that the ONC may
not be in dynamical equilibrium, as the dynamical mass determined from
the kinematic properties of the cluster is twice or more the stellar
mass. Here we follow up on these findings based on our updated
estimates of stellar and gas content in the ONC
(\S\ref{subsection:profile_stars} \&
\ref{subsection:profile_with_gas}).

Proper motions surveys in the ONC date back to the work of
\citep{jones-walker1988}; they measured a 1 dimensional velocity
dispersion $\sigma_v \simeq 2.3$~km~s$^{-1}$. Radial velocity surveys
\citet{sicilia-aguilar2005,furesz2008,tobin2009} derived a nearly
identical velocity dispersion within the ONC region, except for
evidence for lower velocities ($<1.8$~km~s$^{-1}$) for bright members,
and systematic variations with position at scales larger than that
considered in this study, along the north-south filament.

If we consider a singular isothermal profile $\rho(r)=\rho_{0,{\rm
    SIS}}~r^{-2}$, which given the power law exponents 2.2 or 2.07 from
Equations \ref{equation:stellar_profile_fit} and
\ref{equation:stellar_plus_gas_density} is a fair approximation for
the ONC, an average $\sigma_v=2.3$~km~s$^{-1}$, under dynamical
equilibrium would imply a normalization constant
for the density at $r=1$~pc of
\begin{equation}
\rho_{0,{\rm SIS}}=\frac{\sigma_v^2}{2\pi G}= 37\bigg(\frac{\sigma_v}{1~{\rm km~s}^{-1}}\bigg)^2 ~M_\odot~{\rm pc}^{-3} \rightarrow 195~M_\odot~{\rm pc}^{-3}.
\label{equation:IS}
\end{equation}
\noindent
This is about twice the overall value we have estimated from the
contribution of stars and gas in the ONC (Equation
\ref{equation:stellar_plus_gas_density}).
Alternatively, if we consider that the best power law fit of the
estimated stellar plus gas density (Equation
\ref{equation:stellar_plus_gas_density}) has an exponent close to that
of an isothermal sphere, its normalization $\rho_0=100~M_\odot
~{\rm pc}^{-3}$ would lead to a velocity dispersion $\sigma_v \simeq
1.64\:{\rm km\:s^{-1}}$ if in virial equilibrium.

In \S\ref{section:fftime}, below, we find evidence for relatively prolonged star formation history and thus gradual build-up of the ONC, in which case one expects a virialized star cluster to be established before gas removal \citep[e.g.,][]{fellhauer2009}. However, subvirial initial conditions for stellar motions are also a possibility as suggested by studies of dense gas cores \citep[e.g.,]{kirk2007}), with subsequent dynamical evolution investigated by a number of works \citep[e.g.,][]{proszkow2009,allison2009,parker2012}. In this case, the initial density structure would have an even higher normalization than that implied by Equation \ref{equation:IS} (but within the context of a static density structure that does not account for cluster expansion or contraction).

We better characterize the radial dependence of the predicted
$\sigma_v$ from the actual measured density profile of stars alone and
with gas from Equations \ref{equation:stellar_profile_fit},
\ref{equation:gas_density} and
\ref{equation:stellar_plus_gas_density}. For simplicity we assume
isotropic velocities and thus
$\sigma_v = v_{\rm rot}/\sqrt{2}$, which would strictly hold for
a model in equilibrium, and where $v_{\rm rot}$ is the Keplerian
rotational velocity for circular orbits in the potential described by
our volume density profile. The result is shown in the bottom panel of
Figure \ref{figure:plots_profile_with_gas}; we find that given the
actual volume density of stars and gas in the ONC, virial equilibrium
would require $\sigma_v \simeq 1.73$, which is $75\%$ of the measured
velocity dispersion. This indicates that the ONC may be slightly
super-virial, with a virial ratio (kinetic over potential energy)
$q\simeq0.9$;

In this case, the ONC cannot be in perfect dynamical equilibrium, and
should be expanding. This result, however, is affected by some degree
of uncertainty, since it relies on our estimate of total stellar and
gas mass -- which remains a challenging estimate
(\S\ref{subsection:profile_stars}) -- as well as velocity dispersion
measurements from proper motions and/or radial velocity surveys, which in
turn are very sensitive to membership estimates, measurement accuracy
and binary properties of the ONC members.

A current supervirial state of the ONC would be consistent with
general theoretical expectations of dynamical evolution, either
quickly or slowly compared to the dynamical time, from an initially
virialized state as gas is expelled by feedback during the star
cluster formation process. Alternatively, dynamically fast star
cluster formation scenarios can be imagined in which the natal,
transient gas cloud was always in a supervirial state, e.g., if it was
formed or affected by large-scale gas flows \citep{hartmann2012,bonnell2006}.

A supervirial state leading to cluster expansion and dissolution is
also consistent with observed populations of young clusters that
exhibit high ``infant mortality'' with most clusters of a given mass
not surviving at that mass during their first 10~Myr of evolution,
most likely because of their relatively low overall star formation
efficiencies from their natal gas clumps (e.g., Lada \& Lada 2003).

%%%%%%%%%%%%%%%%%%%%%%%%%%%%%%%%%%%%%%%%%%%%%
\begin{figure*}
\center
\includegraphics[width=15cm]{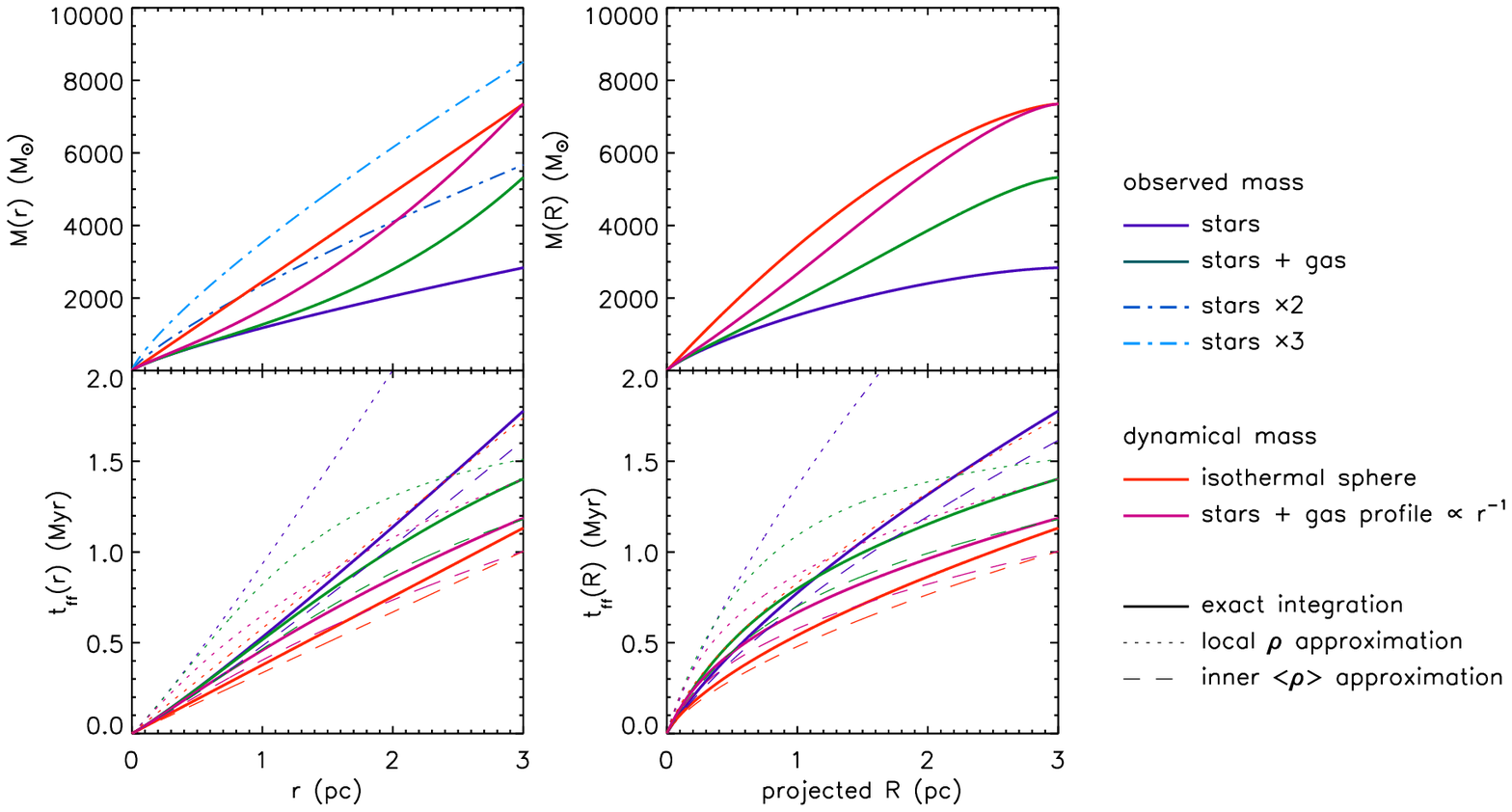}
\caption{
Cumulative mass (\emph{upper panels}) and free-fall time (\emph{bottom
  panels}) as a function of 3D radius, $r$, (\emph{left panels}) and
projected 2D radius, $R$, (\emph{right panels}) for four density distributions:
only the stellar mass profile ($\rho_{\rm gas}=0$) (purple lines); the
stellar plus gas density from Equations
\ref{equation:stellar_profile_fit} and \ref{equation:gas_density}
(green); the isothermal profile that matches the observed $\sigma_v$
in dynamical equilibrium (red); the sum of the present-day density
profile plus a $\rho\propto r^{-1}$ gas profile in order to have a
total mass at $r<3$~pc identical to the previous isothermal model
(magenta). Solid lines are the exact integration from every distance
in the given potential, compared to the approximation $t_{\rm ff}=
[3\pi / (32 G \rho)]^{1/2}$ assuming the local $\rho$ (dotted line) or
the mean $\langle\rho\rangle$ within each radius (dashed
line). \label{figure:fftime}}
\end{figure*}
%%%%%%%%%%%%%%%%%%%%%%%%%%%%%%%%%%%%%%%%%%%%%

%%%%%%%%%%%%%%%%%%%%%%%%%%%%%%%%%%%%%%%%%%%%%
\begin{figure*}
\includegraphics[height=17cm, angle=90]{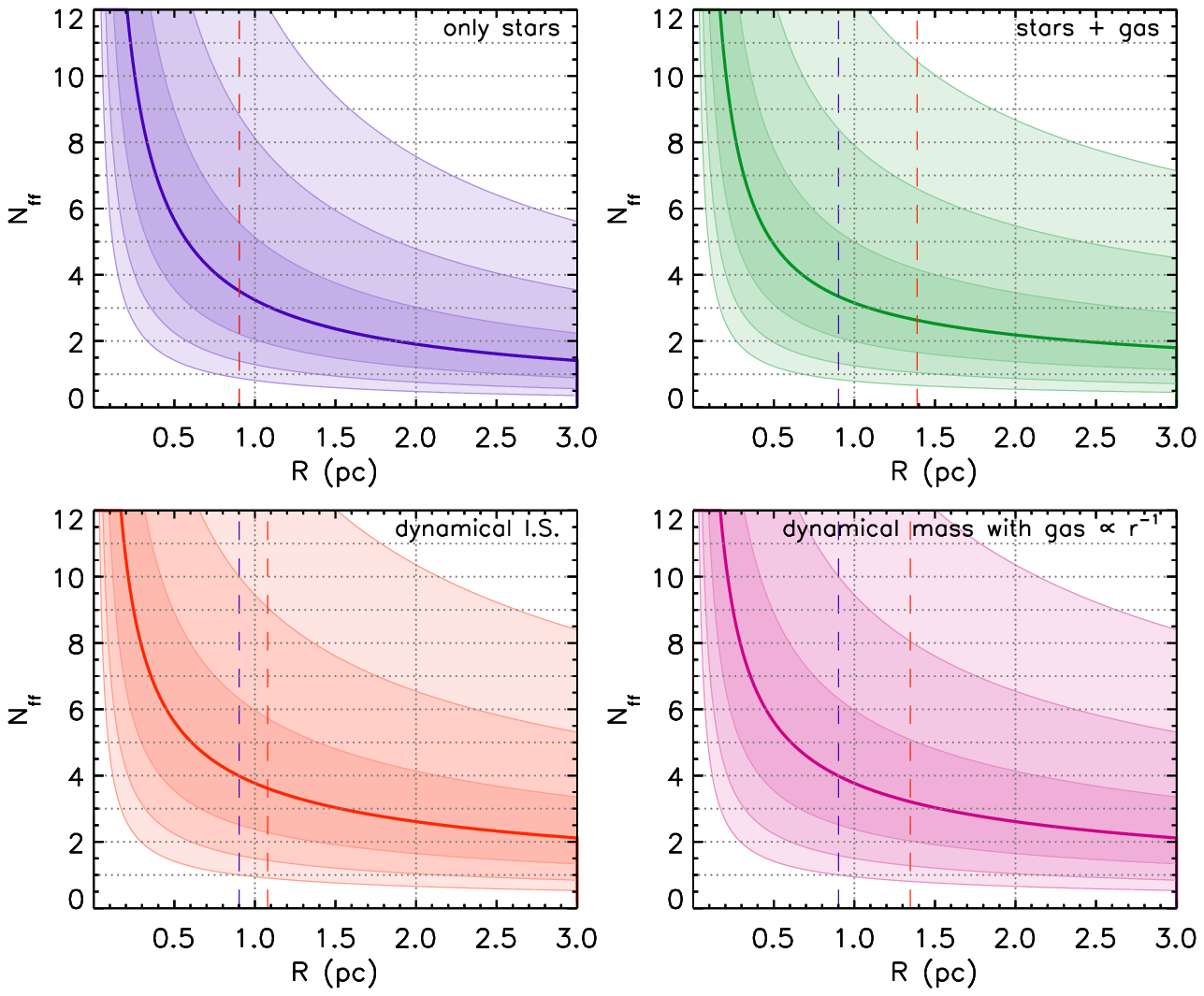}
\caption{
Radial variation (in projected 2D radius, $R$) of the mean age of the
ONC expressed in free-fall times for the four models shown in Figure
\ref{figure:fftime} (thick line). The shaded contours delimit the
$1\sigma$, $2\sigma$ and $3\sigma$ widths of the log-normal age spread
as constrained in the literature (see text). The vertical dashed black
line shows the half-mass radius of the observed stellar density
profile, while the vertical dashed red lines the half-mass radius of
the particular combination of total mass (gas + stars) for each
model. \label{figure:n_fftimes}}
\end{figure*}
%%%%%%%%%%%%%%%%%%%%%%%%%%%%%%%%%%%%%%%%%%%%%

\section{Star Formation Efficiency per Free-Fall Time}
%Free-fall time and age spread}
\label{section:fftime}

A long standing debate in the star formation community concerns the
timescales over which a molecular clump sustains star formation, i.e.,
the duration of star cluster formation. ``Fast'' scenarios
\citep{elmegreen2000,hartmann2001,elmegreen2007,hartmann2007}
predict that star-forming clumps are relatively transient dynamical
entities, with
%rapid dissipation of supersonic turbulence stops the
star cluster formation extending over just one or a few free-fall or
dynamical times. The star formation efficiency per (local) free-fall
time, $\epsilon_{\rm ff}$, would then be relatively high, $\gtrsim
0.1$, depending on the overall star formation efficiency that is
achieved in the forming the cluster from the clump.

Alternatively, in the ``slow'' mode the process is sustained in
quasi-equilibrium for at least several crossing times
\citep[e.g.,][]{tan2006,nakamura2007}, with star formation regulated
by turbulence that is maintained by protostellar outflows or by
support from relatively strong magnetic fields. In these models,
$\epsilon_{\rm ff}$ is relatively low, $\lesssim 0.1$. There is also
more time available for continued accretion of gas to the star-forming
clump from its surroundings.

The extent of the age spread in the ONC (as well as in other clusters)
has been recently constrained by a number of works. The luminosity
spread of its PMS stars, if interpreted as a distribution in radii for
a given mass from a true age spread leads to a large apparent width
($\sigma_{\log t}=0.4$~dex, \citealt{hillenbrand97,dario2010a}) around
a (model dependent) mean age of $\sim 2.5$~Myr. Observational
uncertainties, variability and unresolved binarity cause this age
spread to be overestimated, however. \citet{reggiani2011} showed that
these have a small effect on the overall luminosity
broadening. \citet{jeffries2011} on the other hand posed an upper
limit to the real $\sigma_{\log t}$ of 0.2~dex, from the lack of
correlation between the abundance of circumstellar disks around
members and isochronal ages, suggesting that the apparent luminosity
spread is in large part -- if not all -- due to protostellar accretion
induced changes in the stellar structure evolution
\citep{baraffe2009,baraffe2012}.
However, Hosokawa et al. (2011) found that reasonable levels of
episodic accretion were insufficient to explain the observed
luminosity spread, suggesting significant intrinsic age spreads are
present.
Lastly, \citet{dario2014} excluded a very short age spread, from an
analysis of the bias in the inferred temporal decay of mass accretion
rates induced by uncertain ages of PMS stars, suggesting as a bona
fide compromise of all these independent constraints that there is a
real age spread $\sigma_{\log t}=0.2$~dex around a mean age $t=2.5$~Myr, corresponding to 95\% of
the ONC population with ages between 1 and 6.3~Myr assuming a Gaussian distribution in $\log t$. If instead a uniform distribution
in $\log t$ is assumed, this 95\% of the stars lie in the interval between 1.1 and 5.5~Myr, and for a gaussian distribution in linear age between 0.7 and 4.7~Myr.
We stress that given the amplitude of the apparent age spread compared to the real one, the actual shape of the age distribution is largely unknown; this is also particularly true at very young ages, where the age of individual sources is largely uncertain. Hereafter we will assume a lognormal distribution.

Here we utilize our constrained estimate of the stellar and gas
content of the ONC to translate the age and age distribution of the
ONC in terms of free-fall timescales. We consider
different models for the mass content of the region: first, the
present-day estimates, separately for the radial distribution of mass
volume density of stars alone (Equation
\ref{equation:stellar_profile_fit}) and the sum of stars and gas
(Equation \ref{equation:gas_density}). Second, assuming that the
supervirial state of the ONC we have found in
\S\ref{subsection:equilibrium} is due to recent gas expulsion, we
consider two simple assumptions for the total mass profile before this
event: the singular isothermal sphere that reproduces the observed
velocity dispersion (Equation \ref{equation:IS}) and a model obtained
adding to the measured present-day stellar density profile a gas
profile $\rho_{\rm gas}\propto r^{-1}$ normalized so that the total
mass contained within $r<3$~pc coincides with that of the latter
isothermal sphere.

Figure \ref{figure:fftime} shows the radial dependence of both the
cumulative mass, and the free-fall time $t_{\rm ff}$ for each
model. The solid lines for $t_{\rm ff}$ represent the exact $t_{\rm
  ff}$ calculated by numerical integration of the motion of a test
particle from rest within the modeled potential. For comparison, we
also show the resulting $t_{\rm ff}$ derived adopting the common
approximation valid for a uniform sphere, $t_{\rm ff}= [3\pi / (32 G
  \rho)]^{1/2}$, where we assume for the volume density $\rho$ either
the local one at any given $r$ (dotted lines in Figure
\ref{figure:fftime}) or the average density within the sphere of
radius $r$ (dashed line). The first approximation leads to an
overestimation of $t_{\rm ff}$, since in reality the density increases
moving towards the center. On the other hand the second approximation
leads to results closer to the exact solution, although in this case
$t_{\rm ff}$ are slightly underestimated.  In Figure
\ref{figure:fftime} we also show the mass profile obtained by simply
multiplying the stellar mass by two and three. This shows that the two
models that reproduce the dynamical state in equilibrium are also
compatible with a simple assumptions that the ONC initially had a
similar a similar density profiles as the present-day stellar
distribution, and stars have formed with an efficiency between $\sim
30$ and 50\%. In this case most of the remaining gas has been expelled
by the system during the star cluster formation process.

Using these four models, Figure \ref{figure:n_fftimes} shows the mean
age of the ONC, together with its age spreads in units of 1, 2 and
3$\sigma$ from the mean age, expressed in terms of the number of
free-fall times, $N_{\rm ff}$, in the past at different distances from
the cluster center. To this end we have assumed, as mentioned, a
log-normal age distribution with a width of 0.2~dex around a mean age
of 2.5~Myr. Of course the system becomes increasingly dynamically
older towards the core, even under the assumption of constant average
age and age spread, due to the shorter free fall time at higher
densities.

\begin{figure*}
\plottwo{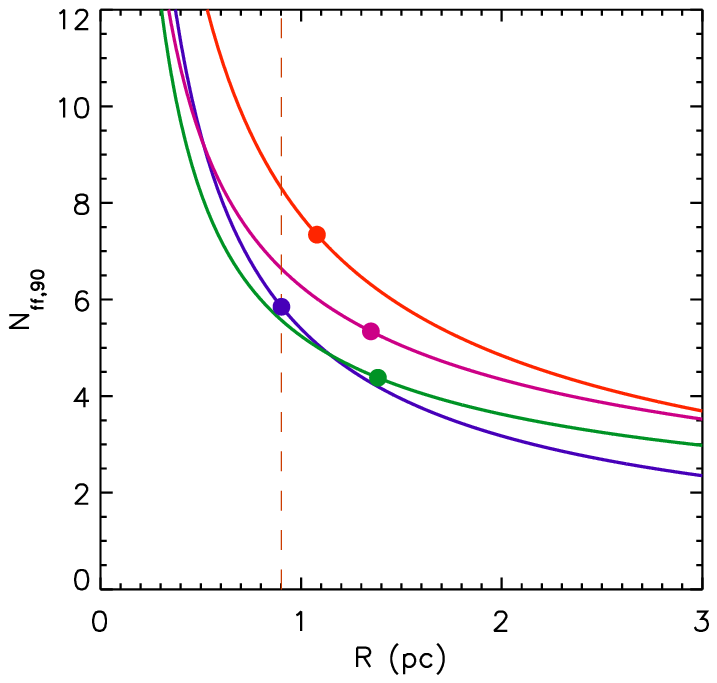}{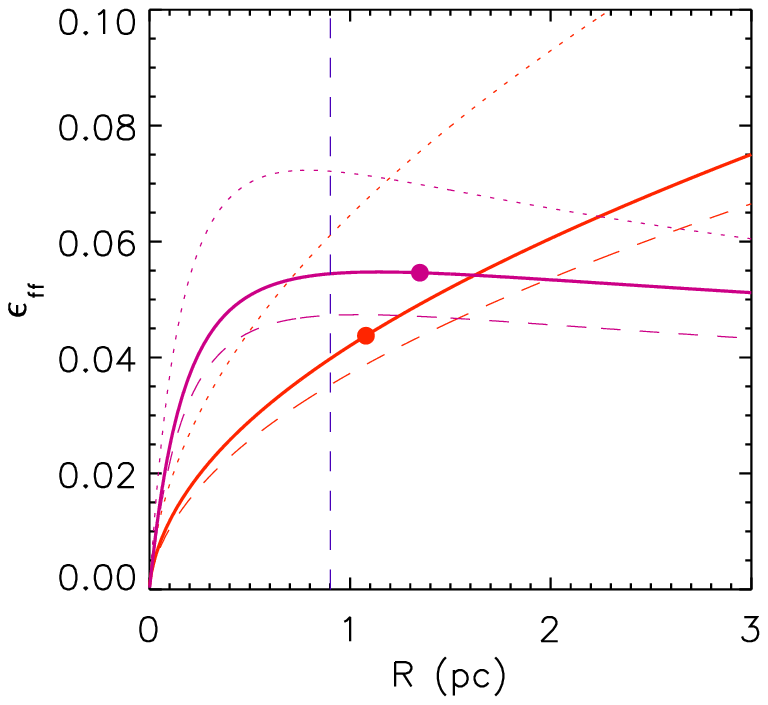}
\caption{
{\em Left panel:} Number of free-fall times needed for the formation
of 90\% of the stellar population in the ONC, as a function of the
projected 2D radius, $R$, from the center, for the four models shown in Figure
\ref{figure:fftime}. The filled circles denote the half-mass radius
each model, and the dashed vertical line the present-day half mass
radius of the stellar density profile. {\em Right panel:} Star
formation efficiency per $t_{\rm ff}$ as a function a radius, for the
two models representing possible initial conditions before gas
removal.\label{figure:nff_90}}
\end{figure*}

Figure \ref{figure:n_fftimes} also highlights
that the age distribution of the ONC spans on average several $t_{\rm
  ff}$ depending on the model. This is clarified in Figure
\ref{figure:nff_90}, left panel; this shows, again as a function of
projected radius from the center, the number $N_{\rm ff,90}$ of free
fall times needed form $90\%$ of the stellar population, in a
symmetric interval with respect to the mean cluster age, for the four
models. When considering the present-day mass distribution, either
from stars alone or from stars and present gas (blue and green) at the
half-mass radius, 90\% of the stellar population has been forming
within 5 to 6 free-fall times. Including the contribution to the
missing mass needed for the ONC to be in dynamical equilibrium, (red
and magenta) increases the number of $t_{\rm ff}$ from 6 to 8 at the
same radii, due to the shorter $t_{\rm ff}$ for these models. In any
case, we find these results to be compatible with a slow star
formation scenario.

If star cluster formation takes place over a relatively extended
period of time, a natural question to ask is whether the
characteristics of star formation, such as the initial mass function,
change systematically during this evolution. Or even more simply, do
massive stars tend to form preferentially near the beginning or end of
star cluster formation? If feedback from massive stars is the primary
agent terminating star cluster formation, then one may expect they
will tend to form near the end of the process. In the ONC, \citet{dario2012} did not find any evidence for a stellar mass versus age
correlation, although such analyses are subject to inherent systematic
uncertainties arising from pre-main sequence stellar evolutionary
models. On the other hand, \citet{getman2014} found the ONC core, where most of the massive stars are located, to be younger than the outskirts.
While massive stars are still forming today in the ONC, such
as ``source I'' in the KL region (see \citealt{tan2014} for a review),
\citep{hoogerwerf2001} have claimed that the four massive ($\sim
20\:M_\odot$) stars $\mu$~Col, AE Aur and the $\iota$~Ori binary formed
in the ONC and were dynamically ejected about 2.5~Myr ago. This
timescale is compatible with the age spread we have adopted from the
analysis of \citet{dario2012} and would indicate that massive star
formation, at least in the case of the ONC, has occurred throughout
the star cluster formation process. It also suggests that the
destructive feedback from massive star formation can be mitigated by
dynamical (self-)ejection of the massive stars---a process likely
enhanced by their migration to the cluster center, as perhaps
exemplified today by the case of $\theta^1$C and its recent
interaction with BN \citep{tan2004,chatterjee2012}.

%%%%%%%%%%%%%%%%%%%%%%%%%%%%%%%%%%%%%%%%%%%%%
%%%%%%%%%%%%%%%%%%%%%%%%%%%%%%%%%%%%%%%%%%%%%

The right panel of Figure \ref{figure:nff_90} shows the star formation
efficiency per free-fall time, $\epsilon_{\rm ff}$. This is simply
estimated as $\epsilon_{\rm ff}=0.9\epsilon_* t_{\rm ff}/t_{\rm
  form,90} = 0.9\epsilon_*/N_{\rm ff,90} $, where $\epsilon_*$ is the
fraction of total mass converted into stars. Since in Figure
\ref{figure:nff_90} we adopt the projected 2D radius, here we assume
$\epsilon_*=\Sigma_*/\Sigma_{\rm tot}$ instead of $\rho_*/\rho_{\rm
  tot}$, as well as the projected $t_{\rm ff}$ as shown in the
bottom-right panel of Figure \ref{figure:fftime}. As in the previous
figures, the red and magenta lines are for the two models we adopt for
the mass content before gas removal, respectively the isothermal
sphere reproducing the observed $\sigma_v$ and the stellar profile
with gas $\sim r^{-1}$. The value of $\epsilon_{\rm ff}$ decreases
towards the core, as a consequence of the smaller $t_{\rm ff}$
compared to the age spread which we assumed radially constant
(consistent with the results of \citealt{dario2012}).

On the other hand, the slow decrease in $\epsilon_{\rm ff}$ at larger
radii for the model with a shallower gas profile is due to the radial
decrease of $\epsilon_*$ since the stellar profile falls more steeply
than the gas in this model. The circles in this figure denote the
value at the half-mass radius, where we find $\epsilon_{\rm ff}\simeq
0.05$. For comparison, in Figure \ref{figure:nff_90} we also show the
same quantity derived from the two approximations of $t_{\rm ff}$ as
computed from the local density at each radius, or the mean density
enclosed within each radius, as in Figure \ref{figure:fftime}.

Our derived values of $\epsilon_{\rm ff}$ are very similar to the
value adopted in the study of Krumholz \& Tan (2007). It is comparable
to the values seen in the simulations of Nakamura \& Li (2007), in
which star formation activity is regulated by protostellar outflow
driven turbulence.

\section{Conclusions}
\label{section:conclusions}

In this work we have reanalyzed the structural and dynamical
properties of the ONC. We based our analysis on a collection of
stellar catalogs, membership estimates and stellar properties from the
latest studies: optically derived stellar parameters, near-infrared
photometry, Spitzer photometry and X-ray data. We have used previous
studies to assess the level of contamination from the Galactic field
in order to constrain the actual stellar population of the
system. Last, we use stellar $A_V$ properties to estimate the ISM
density within the cluster as an additional component adding to the
gravitational potential. Here we briefly summarize our findings.
\begin{enumerate}
\item We determine the center of mass of the ONC, from a subsample of
  sources of known membership. The center is located within the
  Trapezium region, and its position is only weakly sensitive on the
  assumptions for sources without available stellar parameters. We
  also note that the center roughly coincides with the location of the
  point where, according to \citet{chatterjee2012}, the dynamical
  ejection of the BN object from $\theta^1$C took place. $\theta^1$C,
  being the most massive star in the cluster, is expected to migrate
  to this location via dynamical interactions with other cluster
  stars, which thus places a joint constraint on the age of the star
  and the distance of its formation site from the cluster
  center.

\item We analyze the degree of angular substructure of the spatial
  distribution of stars via the angular dispersion parameter,
  $\delta_{\rm ADP,N}$, including its radial dependence.
 A random azimuthal distribution leads to $\delta_{\rm ADP,N}\simeq 1$,
  whereas a degree of additional intrinsic substructure, perhaps
  imparted from initial turbulence in the star-forming gas, increases
  its value. The measured $\delta_{\rm ADP,N}$ in the ONC lies between
  that measured in Globular clusters -- as dynamically old stellar
  systems with no azimuthal substructure -- and Taurus, chosen as an
  example of a very young, dynamically unevolved, clumpy stellar
  system. The dispersion is found to be lower in the core of the ONC
  compared to the outskirts, indicating of higher dynamical processing
  that has erased any initial substructure. However, we also find that
  the elongation of the system along the north-south direction, higher
  at increasing distances from the center, is the major contributor to
  the measured increase of $\delta_{\rm ADP,N}$ with radius, if
  ellipticity is not accounted for. We test the dependence of the
  radial trend of the dispersion on the selection of stellar samples,
  finding no significant variations when different combinations of the
  stellar catalogs are used.
\item We derive the stellar mass surface density and volume density
  profiles of the ONC, for different combinations of the catalogs
  affected by varying degrees of incompleteness and
  contamination. This allows us to accurately extrapolate the bona
  fide profile of the ONC members, which is well reproduced by a
  power-law profile (Equation \ref{equation:stellar_profile_fit}). We
  use measured stellar $A_V$ to derive the average gas density within
  the cluster, which appears to be nearly constant at $\rho_{\rm
    gas}\sim 22\: M_{\odot}\:{\rm pc}^{-3}$, i.e., relatively small
  compared to the stellar density except at large distances from the
  center.
\item We compare the total estimated mass density profile of the ONC
  with literature measurements of the velocity dispersion $\sigma_v$
  in the region, confirming previous claims that the cluster is
  slightly supervirial, indicative that the ONC should be expanding. We expect that this
supervirial state has most likely been caused by relatively recent gas
expulsion, given that the duration of star formation appears to have
been relatively long compared to the free-fall or dynamical time (below).
\item We derive the radial dependence of free-fall time, $t_{\rm ff}$,
  assuming either the present-day measured mass density and different
  simple models for that required by dynamical equilibrium, as
  descriptive of possible configurations before gas disperal. We
  compare $t_{\rm ff}$ with recent constraints on the age and
  intrinsic age spread in the ONC: the cluster appears to be at least
  several $t_{\rm ff}$ old, and $90\%$ of the population has been
  forming over 5 to 8 $t_{\rm ff}$ depending on the assumptions,
  consistent with slow star formation scenarios. From these results we
  infer a star formation efficiency per free-fall time for the
  cluster-forming gas of $\epsilon_{\rm ff}\simeq 0.05$.
\end{enumerate}

\acknowledgments
We thank S.~Chatterjee and A.~Ordonez for their help on simulating clusters for testing the $\delta_{ADP}$ parameter (part of this research will appear in a forthcoming paper). NDR is funded by the Theory Group Fellowship program at UF Astronomy Department.


\begin{thebibliography}{}

\bibitem[Allison et al.(2010)]{allison2010} Allison, R.~J., Goodwin, S.~P., Parker, R.~J., Portegies Zwart, S.~F., \& de Grijs, R.\ 2010, \mnras, 407, 1098
\bibitem[Allison et al.(2009)]{allison2009} Allison, R.~J., Goodwin, S.~P., Parker, R.~J., et al.\ 2009, \apjl, 700, L99
\bibitem[Anderson et al.(2008)]{anderson2008} Anderson, J., Sarajedini, A., Bedin, L.~R., et al.\ 2008, \aj, 135, 2055
\bibitem[Baraffe et al.(2012)]{baraffe2012} Baraffe, I., Vorobyov, E., \& Chabrier, G.\ 2012, \apj, 756, 118
\bibitem[Baraffe et al.(2009)]{baraffe2009} Baraffe, I., Chabrier, G., \& Gallardo, J.\ 2009, \apjl, 702, L27
\bibitem[Bergin et al.(1996)]{bergin1996} Bergin, E.~A., Snell, R.~L., \& Goldsmith, P.~F.\ 1996, \apj, 460, 343
\bibitem[Bertoldi \& McKee(1992)]{bertoldi1992} Bertoldi, F., \& McKee, C.~F.\ 1992, \apj, 395, 140
\bibitem[Bonnell et al.(2006)]{bonnell2006} Bonnell, I.~A., Dobbs, C.~L., Robitaille, T.~P., \& Pringle, J.~E.\ 2006, \mnras, 365, 37
\bibitem[Cartwright \& Whitworth(2004)]{cartwright2004} Cartwright, A., \& Whitworth, A.~P.\ 2004, \mnras, 348, 589
\bibitem[Chatterjee \& Tan(2012)]{chatterjee2012} Chatterjee, S., \& Tan, J.~C.\ 2012, \apj, 754, 152
\bibitem[Da Rio et al.(2014)]{dario2014} Da Rio, N., Jeffries, R.~D., Manara, C.~F., \& Robberto, M.\ 2014, \mnras, 439, 3308
\bibitem[Da Rio et al.(2012)]{dario2012} Da Rio, N., Robberto, M., Hillenbrand, L.~A., Henning, T., \& Stassun, K.~G.\ 2012, \apj, 748, 14
\bibitem[Da Rio et al.(2010a)]{dario2010a} Da Rio, N., Robberto, M., Soderblom, D.~R., et al.\ 2010, \apj, 722, 1092
\bibitem[Da Rio et al.(2010b)]{dario2010b} Da Rio, N., Gouliermis, D.~A., \& Gennaro, M.\ 2010, \apj, 723, 166
\bibitem[Elmegreen(2007)]{elmegreen2007} Elmegreen, B.~G.\ 2007, \apj, 668, 1064
\bibitem[Elmegreen(2000)]{elmegreen2000} Elmegreen, B.~G.\ 2000, \apj, 530, 277
\bibitem[Feigelson et al.(2005)]{feigelson2005} Feigelson, E.~D., Getman, K., Townsley, L., et al.\ 2005, \apjs, 160, 379
\bibitem[F{\H u}r{\'e}sz et al.(2008)]{furesz2008} F{\H u}r{\'e}sz, G., Hartmann, L.~W., Megeath, S.~T., Szentgyorgyi, A.~H., \& Hamden, E.~T.\ 2008, \apj, 676, 1109
\bibitem[Fellhauer et al.(2009)]{fellhauer2009} Fellhauer, M., Wilkinson, M.~I., \& Kroupa, P.\ 2009, \mnras, 397, 954
\bibitem[Getman et al.(2014)]{getman2014} Getman, K.~V., Feigelson, E.~D., \& Kuhn, M.~A.\ 2014, \apj, 787, 109
\bibitem[Getman et al.(2005a)]{getman2005a} Getman, K.~V., Flaccomio, E., Broos, P.~S., et al.\ 2005, \apjs, 160, 319
\bibitem[Getman et al.(2005b)]{getman2005b} Getman, K.~V., Feigelson, E.~D., Grosso, N., et al.\ 2005, \apjs, 160, 353
\bibitem[Grellmann et al.(2013)]{Grellmann2013} Grellmann, R., Preibisch, T., Ratzka, T., et al.\ 2013, \aap, 550, A82
\bibitem[Grosso et al.(2005)]{grosso2005} Grosso, N., Feigelson, E.~D., Getman, K.~V., et al.\ 2005, \apjs, 160, 530
\bibitem[Gutermuth et al.(2009)]{gutermuth2009} Gutermuth, R.~A., Megeath, S.~T., Myers, P.~C., et al.\ 2009, \apjs, 184, 18
\bibitem[Gutermuth et al.(2005)]{gutermuth2005} Gutermuth, R.~A., Megeath, S.~T., Pipher, J.~L., et al.\ 2005, \apj, 632, 397
\bibitem[Hartmann et al.(2012)]{hartmann2012} Hartmann, L., Ballesteros-Paredes, J., \& Heitsch, F.\ 2012, \mnras, 420, 1457
\bibitem[Hartmann \& Burkert(2007)]{hartmann2007} Hartmann, L., \& Burkert, A.\ 2007, \apj, 654, 988
\bibitem[Hartmann et al.(2001)]{hartmann2001} Hartmann, L., Ballesteros-Paredes, J., \& Bergin, E.~A.\ 2001, \apj, 562, 852
\bibitem[Hillenbrand et al.(2013)]{hillenbrad2013} Hillenbrand, L.~A., Hoffer, A.~S., \& Herczeg, G.~J.\ 2013, \aj, 146, 85
\bibitem[Hillenbrand \& Hartmann(1998)]{hillenbrand-hartmann1998} Hillenbrand, L.~A., \& Hartmann, L.~W.\ 1998, \apj, 492, 540
\bibitem[Hillenbrand(1997)]{hillenbrand97} Hillenbrand, L.~A.\ 1997, \aj, 113, 1733
\bibitem[Hoogerwerf et al.(2001)]{hoogerwerf2001} Hoogerwerf, R., de Bruijne, J.~H.~J., \& de Zeeuw, P.~T.\ 2001, \aap, 365, 49
\bibitem[Hosokawa et al.(2011)]{hosokawa2011} Hosokawa, T., Offner, S.~S.~R., \& Krumholz, M.~R.\ 2011, \apj, 738, 140
\bibitem[Huff \& Stahler(2007)]{huff-staher2007} Huff, E.~M., \& Stahler, S.~W.\ 2007, \apj, 666, 281
\bibitem[Jeffries et al.(2011)]{jeffries2011} Jeffries, R.~D., Littlefair, S.~P., Naylor, T., \& Mayne, N.~J.\ 2011, \mnras, 418, 1948
\bibitem[Johnstone \& Bally(1999)]{johnstone1999} Johnstone, D., \& Bally, J.\ 1999, \apjl, 510, L49
\bibitem[Jones \& Walker(1988)]{jones-walker1988} Jones, B.~F., \& Walker, M.~F.\ 1988, \aj, 95, 1755
\bibitem[Kenyon et al.(2008)]{kenyon2008} Kenyon, S.~J., G{\'o}mez, M., \& Whitney, B.~A.\ 2008, Handbook of Star Forming Regions, Volume I, 405
\bibitem[Kirk et al.(2007)]{kirk2007} Kirk, H., Johnstone, D., \& Tafalla, M.\ 2007, \apj, 668, 1042
\bibitem[Kroupa(2001)]{kroupa2001} Kroupa, P.\ 2001, \mnras, 322, 231
\bibitem[Kroupa et al.(2001)]{kroupa2001b} Kroupa, P., Aarseth, S., \& Hurley, J.\ 2001, \mnras, 321, 699
\bibitem[Krumholz \& McKee(2005)]{krumholz2005} Krumholz, M.~R., \& McKee, C.~F.\ 2005, \apj, 630, 250
\bibitem[Kuhn et al.(2014)]{kuhn2014} Kuhn, M.~A., Feigelson, E.~D., Getman, K.~V., et al.\ 2014, \apj, 787, 107
\bibitem[Lada \& Lada(2003)]{ladalada2003} Lada, C.~J., \& Lada, E.~A.\ 2003, \araa, 41, 57
\bibitem[Lombardi et al.(2011)]{lombardi2011} Lombardi, M., Alves, J., \& Lada, C.~J.\ 2011, \aap, 535, A16
\bibitem[Megeath et al.(2012)]{megeath2012} Megeath, S.~T., Gutermuth, R., Muzerolle, J., et al.\ 2012, \aj, 144, 192
\bibitem[Menten et al.(2007)]{menten2007} Menten, K.~M., Reid, M.~J., Forbrich, J., et al.\ 2007, \aap, 474, 515
\bibitem[Muench et al.(2008)]{muench2008} Muench, A., Getman, K., Hillenbrand, L., \& Preibisch, T.\ 2008, Handbook of Star Forming Regions, Volume I, 483
\bibitem[Nakamura \& Li(2014)]{nakamura2014} Nakamura, F., \& Li, Z.-Y.\ 2014, \apj, 783, 115
\bibitem[Nakamura \& Li(2007)]{nakamura2007} Nakamura, F., \& Li, Z.-Y.\ 2007, \apj, 662, 395
\bibitem[O'Dell et al.(2008)]{odell2008} O'Dell, C.~R., Muench, A., Smith, N., \& Zapata, L.\ 2008, Handbook of Star Forming Regions, Volume I, 544
\bibitem[O'dell(2001)]{odell2001} O'dell, C.~R.\ 2001, \araa, 39, 99
\bibitem[Padoan \& Nordlund(2011)]{padoan2011} Padoan, P., \& Nordlund, {\AA}.\ 2011, \apj, 730, 40
\bibitem[Parker et al.(2014)]{parker2014} Parker, R.~J., Wright, N.~J., Goodwin, S.~P., \& Meyer, M.~R.\ 2014, \mnras, 438, 620
\bibitem[Parker \& Meyer(2012)]{parker2012} Parker, R.~J., \& Meyer, M.~R.\ 2012, \mnras, 427, 637
\bibitem[Proszkow \& Adams(2009)]{proszkow2009} Proszkow, E.-M., \& Adams, F.~C.\ 2009, \apjs, 185, 486
\bibitem[Reggiani et al.(2011)]{reggiani2011} Reggiani, M., Robberto, M., Da Rio, N., et al.\ 2011, \aap, 534, A83
\bibitem[Rivilla et al.(2013)]{rivilla2013} Rivilla, V.~M., Mart{\'{\i}}n-Pintado, J., Jim{\'e}nez-Serra, I., \& Rodr{\'{\i}}guez-Franco, A.\ 2013, \aap, 554, A48
\bibitem[Robberto et al.(2010)]{robberto2010} Robberto, M., Soderblom, D.~R., Scandariato, G., et al.\ 2010, \aj, 139, 950
\bibitem[Sarajedini et al.(2007)]{sarajedini2007} Sarajedini, A., Bedin, L.~R., Chaboyer, B., et al.\ 2007, \aj, 133, 1658
\bibitem[Scally et al.(2005)]{scally2005} Scally, A., Clarke, C., \& McCaughrean, M.~J.\ 2005, \mnras, 358, 742
\bibitem[Scandariato et al.(2011)]{scandariato2011} Scandariato, G., Robberto, M., Pagano, I., \& Hillenbrand, L.~A.\ 2011, \aap, 533, A38
\bibitem[Sicilia-Aguilar et al.(2005)]{sicilia-aguilar2005} Sicilia-Aguilar, A., Hartmann, L.~W., Szentgyorgyi, A.~H., et al.\ 2005, \aj, 129, 363
\bibitem[Siess et al.(2000)]{siess2000} Siess, L., Dufour, E., \& Forestini, M.\ 2000, \aap, 358, 593
\bibitem[Skrutskie et al.(2006)]{skrutskie2006} Skrutskie, M.~F., Cutri, R.~M., Stiening, R., et al.\ 2006, \aj, 131, 1163
\bibitem[Tan et al.(2014)]{tan2014} Tan, J.~C., Beltran, M.~T., Caselli, P., et al.\ 2014, arXiv:1402.0919
\bibitem[Tan et al.(2006)]{tan2006} Tan, J.~C., Krumholz, M.~R., \& McKee, C.~F.\ 2006, \apjl, 641, L121
\bibitem[Tan(2004)]{tan2004} Tan, J.~C.\ 2004, \apjl, 607, L47
\bibitem[Tobin et al.(2009)]{tobin2009} Tobin, J.~J., Hartmann, L., Furesz, G., Mateo, M., \& Megeath, S.~T.\ 2009, \apj, 697, 1103
\bibitem[Vuong et al.(2003)]{vuong2003} Vuong, M.~H., Montmerle, T., Grosso, N., et al.\ 2003, \aap, 408, 581
\end{thebibliography}
\end{document}